\documentclass{article} 
\usepackage{iclr2027_conference,times}

\usepackage{amsmath,amsfonts,bm}

\def\eqref#1{equation~\ref{#1}}

\def\1{\bm{1}}

\DeclareMathAlphabet{\mathsfit}{\encodingdefault}{\sfdefault}{m}{sl}
\SetMathAlphabet{\mathsfit}{bold}{\encodingdefault}{\sfdefault}{bx}{n}

\usepackage{hyperref}
\usepackage{makecell}
\usepackage{tabularx}
\usepackage{booktabs}
\usepackage{graphicx}
\usepackage{url}
\usepackage{amsmath,amssymb}
\usepackage{xcolor}
\usepackage[normalem]{ulem} 
\usepackage{pgfplots}
\usepgfplotslibrary{groupplots}
\pgfplotsset{compat=1.17}
\usepackage{braket}

\title{Distance Prediction for Quantum Codes}

\author{\mdseries
\begin{tabular}[t]{@{}l@{}}
\textbf{Nadav Schneider} \\
\texttt{nadavsch@post.bgu.ac.il}
\end{tabular}
\hspace{3em}
\begin{tabular}[t]{@{}l@{}}
\textbf{Eliya Nachmani} \\
\texttt{eliyanac@bgu.ac.il}
\end{tabular}
\AND
\mdseries School of Electrical and Computer Engineering, Ben-Gurion University of the Negev, Israel
}

\iclrfinalcopy

\begin{document}

\maketitle
\lhead{}

\begin{abstract}
The distance of a quantum code is a fundamental metric that dictates its error-correction capability. While quantum low-density parity-check (qLDPC) families such as Bivariate Bicycle (BB) codes offer a promising pathway to low-overhead fault-tolerant quantum computing, computing their distance is challenging. Current exact solvers and even search heuristics that provide a distance upper bound suffer from severe computational bottlenecks, restricting the discovery of optimal codes. In this paper, we construct a dataset of BB codes with corresponding distance upper bounds obtained through high-budget heuristic searches. We introduce BB-Net, an attention-based neural network architecture for predicting the distance of BB codes. It achieves 96.39\% accuracy on the reference distances in the test set, compared with 94.86\% for state-of-the-art heuristics with a budget of up to $10^4$ times more computational runtime. On a subset with certified exact distances, BB-Net achieves 98.94\%. By relieving the distance-calculation bottleneck, BB-Net enables faster screening of large code spaces, particularly in ML-driven search pipelines, and thereby can accelerate the discovery of novel quantum error-correcting codes, an important capability for advancing scalable fault-tolerant quantum computing. Our code is publicly available.
\end{abstract}

\vspace{0.5em}
\hspace{.5em}
\includegraphics[width=1.25em,height=1.15em]{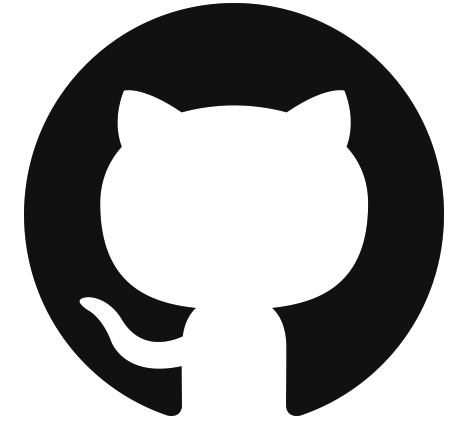}\hspace{.75em}
\parbox{\dimexpr\linewidth-7\fboxsep-7\fboxrule}{\url{https://github.com/NadavSc/BBNet}}
\vspace{-.5em}

\begin{figure}[!h]
    \centering
    \includegraphics[width=0.6\linewidth]{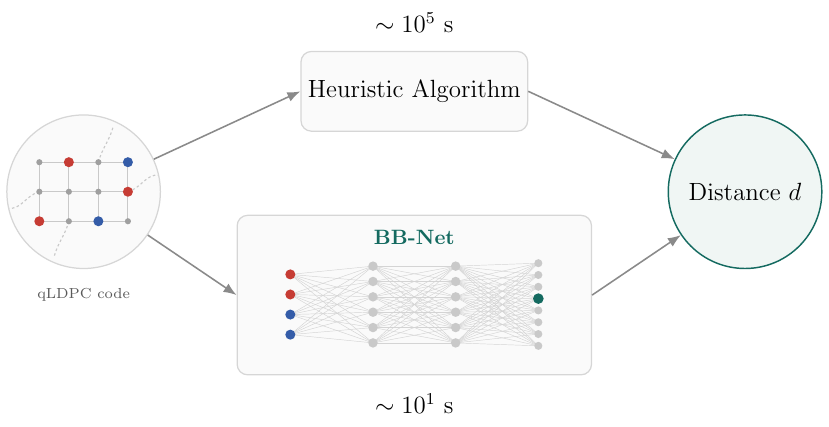}
    \caption{Distance estimation for bivariate bicycle codes through two approaches: computationally intensive heuristic search and orders of magnitude more efficient inference using BB-Net. Numbers refer to a full benchmark evaluation.}
    \label{fig:placeholder}
\end{figure}

\vspace{-15pt}
\section{Introduction}
\vspace{-5pt}
Quantum computing offers the potential to solve computational problems that are intractable for classical methods \citep{steane1998quantum,preskill2012quantum}, with landmark algorithms such as Shor's factoring algorithm \citep{shor1994algorithms} and Grover's search algorithm \citep{grover1996fast} providing prominent examples of quantum computational advantage. Recent experimental advances across multiple quantum-computing platforms have further demonstrated this potential \citep{arute2019quantum,bharti2022noisy,bluvstein2024logical,huang2022quantum,bennett2014quantum,bao2023very}. However, realizing practical, large-scale quantum computation requires overcoming a fundamental challenge: the inherent fragility of physical qubits. Highly susceptible to decoherence and operational errors \citep{burnett2019decoherence,etxezarreta2021time}, physical qubits cannot reliably preserve quantum information for the duration of complex calculations.

Quantum Error Correction (QEC) is therefore indispensable for bridging the gap between noisy physical hardware and reliable computation \citep{shor1995scheme,terhal2015quantum}. Recently, there has been significant interest in the development of quantum Low-Density Parity-Check (qLDPC) codes, which hold the promise of substantially reducing the resource overhead required for fault-tolerant quantum computation \citep{panteleev2021quantum,breuckmann2021quantum}. Within this class, Bivariate Bicycle (BB) codes have emerged as a highly promising family, offering a practical pathway to high-threshold, low-overhead fault tolerance that maps efficiently to physical hardware constraints \citep{bravyi2024high}, and recently, perturbed bivariate bicycle (PBB) codes have been introduced as a non-CSS ansatz that augments the BB polynomial pairs \citep{cruz2026evolutionary}.

The efficacy of any QEC protocol is anchored by a fundamental figure of merit: its minimum distance, $d$. The distance of a QEC code is defined as the minimum number of physical errors required to cause an undetectable logical
failure. It provides a rigorous, decoder-agnostic guarantee of a code's ability to suppress errors. Consequently, discovering codes with optimal distances is a primary objective when designing new fault-tolerant architectures. However, the minimum-distance problem for quantum stabilizer
codes is computationally hard: the corresponding decision problem is NP-complete, and exact distance computation is therefore NP-hard \citep{kapshikar2023hardness, grigorescu2025hardness}. The analogous hardness result for classical linear codes was established by \cite{vardy1997intractability}.

To evaluate code distance, researchers traditionally rely on exact solvers, such as Mixed Integer Linear Programming (MILP) \citep{punekar2010calculating}, or search heuristics such as QDistRnd \citep{pryadko2023qdistrnd}, which provide an upper bound. Exact methods scale exponentially and become computationally intractable even for moderate code sizes. While heuristics offer a relative speedup compared to exact solvers, they remain far from real-time execution. Consequently, evaluating the thousands or millions of candidates encountered in searches over qLDPC code families remains severely constrained by distance-estimation cost \citep{webster2026distance}. A common strategy is therefore to perform distance screening: limited computational budget is first used to filter candidates, after which computationally intensive heuristics or exact solvers are applied only to the most promising codes to tighten bounds or obtain certificates \citep{qian2026multi,yan2026omniqec,cruz2026evolutionary}. However, the effectiveness of such pipelines remains limited by the runtime of existing distance-estimation methods, leaving large portions of the design space unexplored.

In this paper, we address this critical bottleneck by introducing BB-Net, an attention-based neural network architecture for both BB and PBB codes. We propose a compact tokenization scheme that exploits the algebraic structure of BB codes, enabling BB-Net to learn informative latent representations for distance prediction. By relieving this bottleneck, BB-Net enables fast evaluation of large code spaces, particularly in ML-driven search pipelines, and thereby can accelerate the discovery of novel QEC codes, an important capability for advancing scalable fault-tolerant quantum computing.

\textbf{Our main contributions are summarized as follows:}
\begin{itemize}
    \item We frame distance estimation for BB codes as a machine-learning problem: a network is trained on labels from expensive heuristic searches and predicts them at a small fraction of the cost. To our knowledge, this is the first work to predict quantum-code distance directly with a learned model.
    \item We present a designated dataset of approximately 58,000 code examples of both CSS BB and non-CSS PBB types with a corresponding distance upper bound.
    \item We propose a highly compressed input representation for the codes and an attention-based sequence neural network architecture, BB-Net, for this task.
    \item We show that BB-Net's accuracy on the test benchmark is 96.39\% compared to 94.86\% by state-of-the-art heuristics with a budget of up to $10^4$ more computational runtime, and it achieves 98.94\% on certified exact distance subset test codes.
\end{itemize}

\newpage
\vspace{-10pt}
\section{Related Work}
\vspace{-10pt}
Quantum-code distance can be computed or bounded by a variety of algebraic, combinatorial, optimization, and randomized techniques. Exact
methods search systematically enough to certify that no logical operator exists below a given weight \citep{dumer2017distance}. These include codeword-enumeration methods such as Brouwer--Zimmermann \citep{grassl2006searching,hernando2026fast},
error-enumeration methods such as bipartition matching and linked-cluster search \citep{dumer2014numerical}, and solver-based formulations using mixed-integer programming \citep{bestuzheva2021scip} or SAT/MaxSAT
\citep{chen2026sat}. Heuristic methods can provide a powerful alternative and substantially faster than the exact methods, however, they only satisfy an upper bound to the distance. The methods include random-information-set such as QDistRnd \citep{pryadko2023qdistrnd}, that uses random information-set sampling and Gaussian elimination to search for low-weight logical operators, evolutionary methods such as QDistEvol \citep{webster2025engineering} that replaces random sampling with an evolutionary search over column permutations, syndrome-decoder-based distance searches using BP-OSD \citep{roffe2020decoding}, and Stim-based methods such as GEStim, CCStim, and UEStim that search for undetectable logical errors, targeting increasingly general error structures and detector degrees \citep{gidney2021stim}.

Most machine-learning work on error-correcting codes addresses decoding: a model observes a noisy word or syndrome and predicts an error or corrected codeword \cite{choukroun2022denoising,krenn2023artificial,wang2024artificial,demarti2024decoding,varsamopoulos2018decoding,varsamopoulos2019comparing,liu2019neural,harper2020efficient,maskara2019advantages,meinerz2022scalable}. Some decoders are based on reinforcement learning \cite{colomer2020reinforcement,sweke2021reinforcement,fitzek2020deep,veeresh2024rl,andreasson2019quantum}, others on GNN \citep{lange2025data}, transformers \citep{choukroun2024foundation,bausch2024learning,senior2025scalable,park2025crossmpt,zenati2026saq,peled2026neural,blue2026machine}. Beyond decoding, recent works have been discovering new practical QEC codes with LLMs \citep{qian2026multi,yan2026omniqec,cruz2026evolutionary}.

\vspace{-5pt}
\section{Background}
\vspace{-5pt}
\subsection{Quantum Error-Correcting Codes}
\vspace{-5pt}

A classical linear code encodes $k$ information bits into $n$ binary symbols by imposing parity constraints through a binary matrix $H$ \citep{hamming1950error}. Valid codewords satisfy $Hx^\top=0$, while an error $e$ produces the syndrome $s=He^\top$. Quantum error correction follows the same syndrome-based principle but within a two-level systems described by a state vector $\ket{\psi}$ \citep{shor1995scheme,wootters1982single}, as a qubit can exist in coherent superpositions:
\begin{equation}
    \ket{\psi}
    = \alpha\ket{0}+\beta\ket{1},
    \qquad
    \text{where }
    \alpha,\beta\in\mathbb{C},
    \quad
    \lvert\alpha\rvert^2+\lvert\beta\rvert^2=1.
\end{equation}
Quantum errors are expressed using the Pauli basis $\{I,X,Y,Z\}$, representing identity, bit-flip, combined bit-and-phase-flip, and phase-flip operations. An $n$-qubit Pauli error is a tensor product of these operators, and its weight is the number of qubits on which it acts
nontrivially.

The stabilizer formalism \citep{gottesman1997stabilizer} provides a
convenient algebraic description of quantum codes. An $[[n,k,d]]$ stabilizer code is defined by a commuting set of Pauli operators within the Hilbert space $\mathcal{H}_2^{n}$ whose common $+1$ eigenspace forms the codespace.
The stabilizers generate an Abelian group $\mathcal{S}$ that does not contain $-I$. A group with $n-k$ independent generators defines a $2^k$-dimensional codespace and therefore encodes $k$ logical qubits. Measuring the generators produces a binary syndrome indicating which stabilizers anticommute with an error, without revealing the encoded logical state. In CSS codes \citep{calderbank1996good,steane1996multiple,steane1996error}, the stabilizers are separated into $X$ and $Z$ type checks described by binary matrices $H_X$
and $H_Z$, which satisfy
\begin{equation}
    H_XH_Z^\top=0 \pmod 2.
\end{equation}
This condition ensures that all $X$ and $Z$ type
stabilizers commute. For an error represented by binary $X$ and $Z$
components $(e_X,e_Z)$, the corresponding syndromes are
$s_X=H_Xe_Z^\top$ and $s_Z=H_Ze_X^\top$.

\newpage
\subsection{Quantum Code Distance Finding}
\label{sec:distance_finding}
\vspace{-5pt}
Computing the distance of a quantum code amounts to finding the minimum-weight nontrivial logical operator. Let $\mathcal{N}(\mathcal{S})$ denote the set of Pauli operators that commute with every element of the stabilizer group $\mathcal{S}$. The distance is formally
\begin{equation}
    d
    =
    \min_{L\in
    \mathcal{N}(\mathcal{S})\setminus\mathcal{S}}
    \operatorname{wt}(L).
\end{equation}
where $\operatorname{wt}(L)$ denotes the Pauli weight of $L$, i.e., the number of physical qubits on which $L$ acts nontrivially with an $X$, $Y$, or $Z$ operator rather than the identity. Thus, a logical operator preserves the codespace but acts nontrivially on the encoded information. A distance-$d$ code detects every error of weight less than $d$ and corrects every error of weight at most $\lfloor(d-1)/2\rfloor$. For a CSS code, the $X$ and $Z$ distances can be computed independently,
with the overall distance given by $d=\min(d_X,d_Z)$. The decision version of the problem asks whether a nontrivial logical operator of weight at most a given threshold exists. This problem is NP-complete for general stabilizer codes, and the exact distance computation is therefore NP-hard \citep{kapshikar2023hardness}. Distance-finding algorithms broadly trade computational cost for the strength of the guarantee they provide \citep{webster2026distance}. A completed exact search certifies the distance, whereas an incomplete or heuristic search generally provides only a bound. During the search, such algorithms may maintain both a lower bound
$d_{\mathrm{lower}}$ and an upper bound $d_{\mathrm{upper}}$.
When $d_{\mathrm{lower}}=d_{\mathrm{upper}}$, the distance is certified exactly.
However, exact methods may require exhaustive runtimes and are frequently terminated before convergence on larger codes, in which case only partial bounds are available. Heuristic methods instead search for low-weight logical operators without certifying that no lower-weight operator exists. For a CSS code with parity-check matrices $H_X$ and $H_Z$, a candidate
$Z$-type witness $\mathbf{v}\in\mathbb{F}_2^n$ is accepted only if
\begin{equation}
    H_X\mathbf{v}^{\top}=0,
    \qquad
    \mathbf{v}\notin\operatorname{rowspan}(H_Z),
    \qquad
    \operatorname{wt}(\mathbf{v})=\hat d,
\end{equation}
and analogously an $X$-type witness must satisfy the corresponding
conditions with $H_X$ and $H_Z$ interchanged. A heuristic result is considered invalid if any of these conditions fail.

\vspace{-5pt}
\subsection{Bivariate Bicycle Codes}
\label{sec:bb_background}
\vspace{-5pt}
Bivariate Bicycle (BB) codes are a family of quantum low-density parity-check (qLDPC) codes introduced in their present form by \citet{bravyi2024high}. They use a two-dimensional cyclic structure and provide a compact algebraic description through sparse bivariate polynomials. BB codes are attractive because they combine sparse constant-weight stabilizer checks with comparatively high encoding rates, making them promising candidates for low-overhead fault-tolerant quantum memories.

A BB code is defined over the finite Abelian group
$\mathbb{Z}_{\ell}\times\mathbb{Z}_{m}$ and can be represented using two sparse polynomials
\begin{equation}
    A(x,y)=\sum_{(a,b)\in S_A}x^ay^b,
    \qquad
    B(x,y)=\sum_{(a,b)\in S_B}x^ay^b,
\end{equation}
where $S_A$ and $S_B$ are sets of exponent pairs. The periodic relations $x^\ell=1$ and $y^m=1$ make the polynomial coordinates cyclic. Each monomial $x^ay^b$ corresponds to a two-dimensional cyclic shift, and the resulting $\ell m\times \ell m$ binary matrices $A$ and $B$ define the CSS parity check matrices
\begin{equation}
    H_X=[A B],
    \qquad
    H_Z=[B^\top A^\top].
\end{equation}
The code therefore has block length $n=2\ell m$. The construction typically use sparse weight-three polynomials, so each stabilizer has constant weight six. This gives BB codes their LDPC structure while retaining a highly regular algebraic organization. Different choices of $(\ell,m)$ and of the polynomial supports $S_A$ and $S_B$ generate different codes with different values of $k$ and $d$, creating a large design space that must be searched efficiently. Our study also includes non-CSS Perturbed Bivariate Bicycle (PBB) \citep{cruz2026evolutionary} codes, which extend this construction using four sparse bivariate polynomials $(A,B,C,D)$ and do not admit the same separation into purely $X$ and $Z$ type parity-check matrices:
\begin{equation}
    H
    =
    \left[
    \begin{array}{cccc}
        A & B & C          & D          \\
        0 & 0 & B^{\top}  & A^{\top}
    \end{array}
    \right],
    \label{eq:pbb_stabilizer}
\end{equation}
where the first stabilizer has both \(X\)-type (\(A\),\(B\)) and \(Z\)-type (\(C\),\(D\)) components and the second is purely \(Z\)-type.

\vspace{-5pt}
\section{Method}
\label{sec:method}
\begin{figure}
    \centering
    \includegraphics[width=0.8\linewidth]{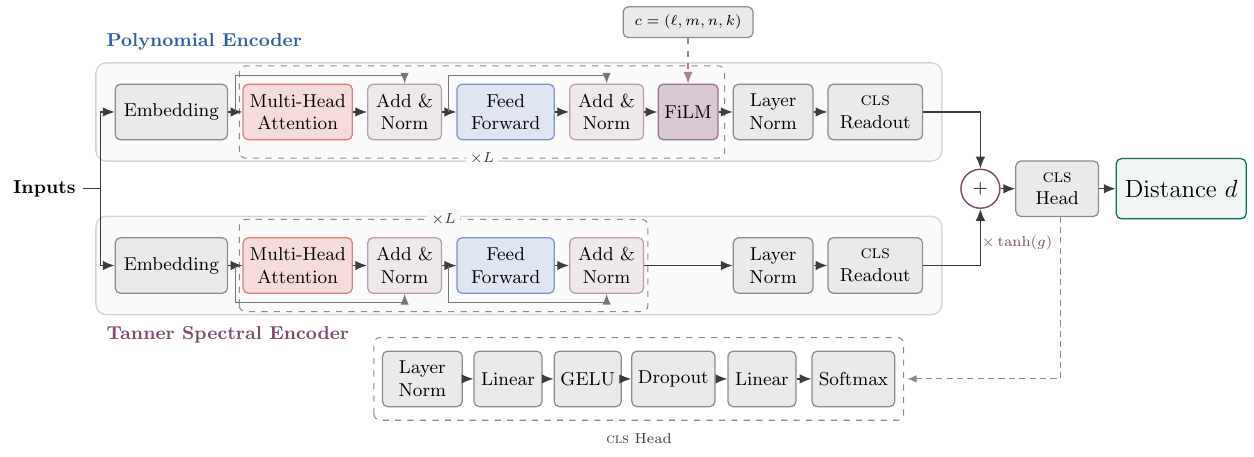}
    \caption{\textbf{BB-Net architecture}. A BB code is processed by two encoders; The polynomial encoder applies self-attention to monomial tokens and is conditioned on the code parameters through FiLM layers. In parallel, the Tanner spectral encoder processes Fourier-character tokens using attention as well. Each branch produces a code latent representation, which is combined through a learned gated residual connection. The fused representation then passed to a classification head.}
    \vspace{-10pt}
    \label{fig:bb_net_architecture}
\end{figure}

\vspace{-5pt}
We propose BB-Net, an attention-based sequence architecture that consists of two complementary encoding branches: a polynomial encoder, which captures the local algebraic structure of the code specification, and a Tanner spectral encoder, which captures global structural properties of the corresponding Tanner graph. The code is first tokenized into a compact cyclic representation that preserves its underlying algebraic structure while avoiding the need to explicitly construct large parity-check matrices, and then proceeds through an embedding block. The latent representations produced by the two encoders are subsequently combined through a gate and passed to a distance-classification head, which predicts the code distance. Architecture is described in \autoref{fig:bb_net_architecture}.

\subsection{Code Tokenization and Embedding}
\label{sec:tokenization}
\begin{figure}
    \centering
    \includegraphics[width=0.7\linewidth]{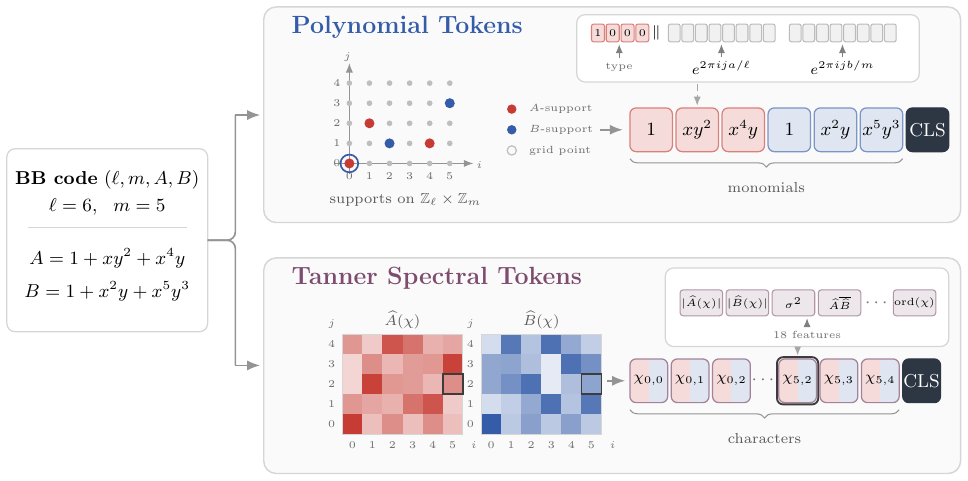}
    \caption{\textbf{BB-Net tokenization and embedding.} A BB code specified by sparse polynomials \(A\) and \(B\) is mapped to two complementary representations: (i) polynomial tokens encode monomial supports on \(\mathbb{Z}_\ell \times \mathbb{Z}_m\) with the type (A,B) indicator as a one-hot vector and the first eight Fourier harmonics, represented by sine and cosine components along each lattice dimension, yielding a 32-dimensional positional representation. (ii) Tanner spectral tokens encode the responses of the full polynomials at each group character \(\chi\) using 18-dimensional character features. Learned CLS tokens are used as latent vector aggregators for the downstream classification task.}
    \vspace{-10pt}
    \label{fig:tokenize}
\end{figure}

\vspace{-5pt}
BB-Net constructs two complementary token representations from the sparse polynomial specification of each code: (i) polynomial tokens, which describe the local algebraic relations between monomials, and (ii) Tanner spectral tokens, which provide a global representation of the corresponding cyclic Tanner structure. Both representations are generated directly from the polynomial supports, without expanding the code into its full parity-check matrices. These token representations are subsequently mapped to continuous feature spaces by the encoder's embedding block, and learned CLS tokens are used as latent vector aggregators for the downstream classification task (\autoref{fig:tokenize}).

\paragraph{Polynomial tokens.}
Given the support sets $S_A,S_B$ (and $S_C,S_D$ for PBB codes), we first
create one monomial token for every exponent pair appearing in the code.
Each monomial token is associated with its polynomial identity,
\begin{equation}
    \mathcal{T}_{\mathrm{monomial}}
    =
    \{(P,a,b):(a,b)\in S_P,\;
    P\in\{A,B,C,D\}\}.
\end{equation}

For a monomial token $(P,a,b)$, the component identity $P\in\{A,B,C,D\}$ is embedded by a one-hot vector. The a and b components are embedded as cyclic coordinates without imposing a linear ordering on the torus. For harmonic index $r=1,\ldots,8$, we compute
\begin{equation}
\boldsymbol{\phi}(a,b)=
\left[
\sin\!\left(\frac{2\pi r a}{\ell}\right),
\cos\!\left(\frac{2\pi r a}{\ell}\right),
\sin\!\left(\frac{2\pi r b}{m}\right),
\cos\!\left(\frac{2\pi r b}{m}\right)
\right]_{r=1}^{8}\in\mathbb{R}^{32}.
\end{equation}
This allows the network to distinguish identical cyclic coordinates appearing in different components of the code \citep{tancik2020fourier}.

\paragraph{Tanner spectral tokens.}
The second representation exploits the translational symmetry of the BB Tanner graph. Tanner graphs provide the standard bipartite representation of sparse parity constraints \citep{tanner1981recursive}; for BB codes, the additional cyclic structure makes it possible to work with the associated group lift rather than explicitly materializing all Tanner-graph nodes and edges. Fourier-domain representations have similarly been used to analyze the structure of quasi-cyclic LDPC codes constructed from circulant matrices \citep{diao2012matrix}, while the spectra of group-based graph lifts can be decomposed through representations of the underlying
group \citep{agarwal2019expansion}. Specifically, BB-Net represents the cyclic lift in the character domain.
Since $G=\mathbb{Z}_{\ell}\times\mathbb{Z}_{m}$ is finite and Abelian, its dual group contains $\ell m$ one-dimensional
characters. For $(u,v)\in
\mathbb{Z}_{\ell}\times\mathbb{Z}_{m}$,
\begin{equation}
    \chi_{u,v}(a,b)
    =
    \exp\left[
        -2\pi i
        \left(
            \frac{ua}{\ell}
            +
            \frac{vb}{m}
        \right)
    \right].
\end{equation}

We construct one spectral token for every character $\chi_{u,v}$. For each character, the sparse polynomial supports are evaluated as
\vspace{-2pt}
\begin{equation}
    \widehat{P}(\chi_{u,v})
    =
    \sum_{(a,b)\in S_P}
    \chi_{u,v}(a,b),
    \qquad
    P\in\{A,B,C,D\}.
\end{equation}
Thus, each character indexes a particular Fourier mode of the cyclic lift, and the corresponding responses summarize how the local polynomial connections combine at that mode. Each character $\chi$ is embedded by a compact $18$-dimensional feature vector whose entries are derived from the responses evaluated at that same character. These include normalized polynomial magnitudes such as $|\widehat{A}(\chi)|$ and $|\widehat{B}(\chi)|$, spectral-energy terms, cross-polynomial interactions such as
$\widehat{A}(\chi)\widehat{B}(\chi)^{*}$, relative spectral differences, and invariant properties of the character itself. The spectral sequence contains $|G|=\ell m=\frac{n}{2}$ tokens, including the trivial character. A character is therefore identified by its cyclic invariants and by the responses of the complete polynomial specification at that character, rather than by an arbitrary flattened Tanner-node index.

\vspace{-7pt}
\subsection{Architecture}
\label{sec:encoders}
\vspace{-5pt}
Each branch uses a stack of L blocks of multi-head self-attention \citep{vaswani2017attention} and feed-forward with residual connections and layer normalization. The polynomial branch is additionally conditioned on the code's information $\mathbf{c}=(\ell,m,n,k)$ with FiLM \citep{perez2018film}, an MLP that maps $\mathbf{c}$ to feature-wise scale and shift vectors. Through self-attention, the branch-specific \textsc{cls} tokens aggregate the
two token sequences into latent vectors,
$\mathbf{h}_{\mathrm{polynomial}},
\mathbf{h}_{\mathrm{tanner}}
\in \mathbb{R}^{d_{\mathrm{model}}},$
which are subsequently combined through a learned residual gate: $\mathbf{h}_{\mathrm{fuse}}=\mathbf{h}_{\mathrm{polynomial}}+\tanh(\mathbf{g})\odot \mathbf{h}_{\mathrm{tanner}}$,
where $\mathbf{g}$ is learned jointly with the network. This parameterization retains a direct polynomial path while allowing the model to use global spectral information. 

The classification head follows layer normalization with a linear layer, GELU activation, dropout, another linear projection, and a softmax. Let \(\mathcal{D}=\{d_{\min},\ldots,d_{\max}\}\) denote the effective classes, the output layer's length is $d_{\max}-d_{\min}+1$ and the model is trained with a categorical cross-entropy over these classes.

\vspace{-7pt}
\subsection{Canonicalization and Training Augmentation}
\label{sec:invariances}
\vspace{-5pt}
\paragraph{\textbf{Canonicalization.}}  Shift canonicalization maps each input to a deterministic representative of its cyclic-shift equivalence class.
Let
\(S_P\subseteq\mathbb{Z}_{\ell}\times\mathbb{Z}_{m}\) be the support of a
polynomial and let \(\mathbf{s}=(s_1,s_2)\). We define the translated support
as
\begin{equation}
\tau_{\mathbf{s}}(S_P)
=
\left\{
\left(
(a-s_1)\bmod\ell,\,
(b-s_2)\bmod m
\right)
:
(a,b)\in S_P
\right\}.
\end{equation}
This transformation is equivalent to multiplying the corresponding polynomial by the monomial \(x^{-s_1}y^{-s_2}\), with the exponents evaluated
modulo \((\ell,m)\), thus, inducing a cyclic row or column permutations and therefore does
not change the code parameters, including its distance. For each polynomial, every monomial is used in turn as an anchor, and the support is cyclically shifted so that the anchor moves to the origin \((0,0)\). The resulting exponent pairs are sorted, and the lexicographically smallest representation is selected as the canonical form \citep{feulner2009automorphism}. For PBB codes we apply a single translation to all polynomials.

\paragraph{\textbf{Augmentation.}}
Each code is augmented using randomly sampled
distance-preserving symmetries of its algebraic representation \citep{chen2020group}.
Let $G=\mathbb{Z}_{\ell}\times\mathbb{Z}_{m}.$
For a CSS BB code, we sample an automorphism $\varphi\in\operatorname{Aut}(G)$ \citep{hillar2007automorphisms}
and apply it simultaneously to the exponent coordinates of both polynomial supports. Specifically, each monomial \(x^a y^b\) is mapped to \(x^{a'}y^{b'}\), where $(a',b')=\varphi(a,b).$ In addition, independent uniformly sampled cyclic translations are applied
to \(S_A\) and \(S_B\), and the two polynomial registers may be interchanged, that is, $(A,B)\longmapsto(B,A).$ For a non-CSS PBB code, the same sampled automorphism is applied simultaneously to all four polynomial supports. The register interchange is $(A,B,C,D)\longmapsto(B,A,D,C),$ and a single uniformly sampled cyclic translation is applied simultaneously
to \(S_A,S_B,S_C,\) and \(S_D\).

\vspace{-7pt}
\section{Dataset}
\label{sec:dataset}
\vspace{-7pt}
We construct the corpus from CSS BB codes and non-CSS PBB extensions. We restrict the corpus to $n=2\ell m\leq360$ with $\ell,m\leq30$. Candidate generation covers separable, coordinate-swapped, mixed-monomial, trinomial, and adversarial BB specifications together with valid PBB expansions. Then,
candidate specifications are reduced modulo $(\ell,m)$, with coincident monomials canceled over $\mathbb{F}_2$. We keep only candidates with at least two surviving terms in $A$, $B$. for PBB codes, $C$ and $D$ must remain nonempty. We then exclude codes with zero logical dimension, $k= 0$. For PBB candidates, we further require the stabilizer generators to commute, verified through the symplectic commutation condition $H_XH_Z^{\top}+H_ZH_X^{\top}=0 \pmod 2$.
Finally, duplicate specifications are removed, including symmetry-equivalent.

To find the distance of each code, we apply the BP-OSD, QDistRnd, and QDistEvol heuristics, using a large computational budget of 10,000 trials per code. The smallest distance identified across the three methods is considered the target distance. Since these heuristic methods do not provide certificates, the resulting values are upper bounds on the true code distance. The final dataset contains 58,704 examples with upper-bound distances ranging from 2 to 25, and is partitioned into training, validation and test sets using a 80:10:10 split. We ran the exact solver MILP \citep{punekar2010calculating} on all 5,870 test codes with a time limit of 2 hours per code; it certified the distance of 2,366 of them. On every certified code, the 10,000-trial heuristic label equals the certified distance.

\vspace{-7pt}
\section{Experimental Results}
\label{sec:experimental_resukts}
\vspace{-5pt}
To evaluate BB-Net, we compare its performance against six distance-finding heuristic methods; QDistRnd \citep{pryadko2023qdistrnd}, QDistEvol \citep{webster2025engineering}, BP-OSD \citep{roffe2020decoding}, GEStim, CCStim, and UEStim \citep{gidney2021stim}, with a budget of up to $10^4$ times more computational runtime for the benchmark (\autoref{sec:performance_analysis}). Then, we evaluate the methods on certified and high upper bound codes (\autoref{sec:certified_and_high}). AUC for distance screening is reported (\autoref{sec:distance_screening}), and finally, we conduct ablation experiments to justify our architecture design (\autoref{sec:ablation}). Experiments' configurations as well as more experimental analysis are in \autoref{app:experimental_results}.

\vspace{-7pt}
\subsection{Test Benchmark}
\label{sec:performance_analysis}
\vspace{-5pt}
Table~\ref{tab:benchmark_200} presents results on the test benchmark. To reliably identify the correct distance regime, which is particularly relevant for comparing candidates' codes, accuracy is measured using a tolerance of $\pm 2$ around the reference distance. BB-Net achieves the highest accuracy for the referenced distances, reaching 96.39\% with the lowest RMSE of 1.131. The heuristics methods require repeated stochastic searches, denoted as trials per code. BB-Net evaluates a code in 0.005\,s, compared with 1.0\,s for QDistEvol, 3.3\,s for QDistRnd, and 19.1\,s for BP-OSD. The Stim-based methods exhibit substantially lower accuracy and, in the case of GEStim, rarely return a solution. Overall, the results demonstrate that BB-Net achieves high accuracy with substantially better efficiency, enabling near-real-time distance estimation.

\vspace{-5pt}
\begin{table}[!h]
\centering
\caption{Test benchmark across common heuristic methods (5,870 codes)}
\label{tab:benchmark_200}
\small
\setlength{\tabcolsep}{3pt}
\begin{tabular}{l rrrrrr | r}
\toprule
& QDistRnd & QDistEvol & BP-OSD & GEStim & CCStim & UEStim & BB-Net (ours) \\
\midrule
Trials Per Code               & 200 & 200 & 200 & 1 & 1 & 3 & 1  \\
Time Per Code [s] ($\downarrow$)    & 3.3 & 1.0 & 19.1 & 1.2 & 1.2 & 13.1 & \textbf{0.005} \\
Total Time [s] ($\downarrow$)       & 19,383 & 6,057 & 112,275 & 6,940 & 7,164 & 76,775 & \textbf{26.8} \\
RMSE ($\downarrow$)                 & 2.319 & 1.485 & 23.322 & 4.950 &  9.309 & 1.933 & \textbf{1.131} \\
Accuracy ($\uparrow$)           & 92.20\% & 94.86\% & 82.23\% & 0.02\% & 34.60\% & 61.23\% & \textbf{96.39\%} \\
\bottomrule
\end{tabular}
\end{table}
\subsection{Certified and High Upper Bound Codes}
\label{sec:certified_and_high}
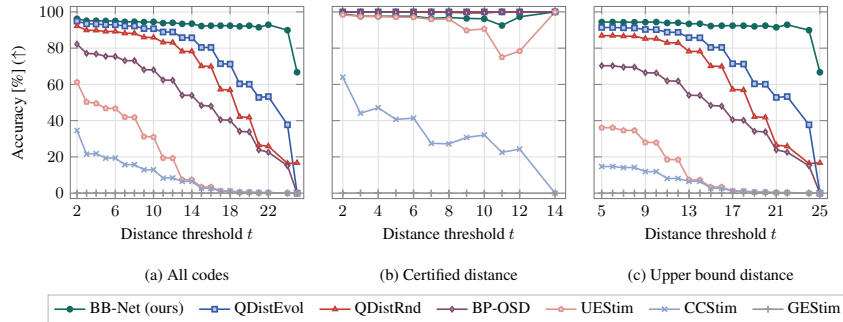
\begin{figure}[!h]
\centering
\definecolor{Ared} {RGB}{198, 60, 52}%
\definecolor{Bblue}{RGB}{ 52, 92,168}%
\definecolor{specc}{RGB}{ 20,106, 96}%
\definecolor{dark} {RGB}{ 60, 60, 60}%
\definecolor{greyc}{RGB}{150,150,150}%
\colorlet{mixc}{Ared!50!Bblue}%
\resizebox{0.8\textwidth}{!}{%
\begin{tikzpicture}
\begin{groupplot}[
  group style={group size=3 by 1, horizontal sep=0.55cm,
               y descriptions at=edge left},
  width=6.1cm, height=5.3cm, clip=false,
  ymin=-3, ymax=103, ytick={0,20,40,60,80,100},
  ylabel={Accuracy [\%]  ($\uparrow$)},
  xlabel={Distance threshold $t$},
  xlabel style={font=\small}, ylabel style={font=\small},
  tick label style={font=\footnotesize},
  grid=major, grid style={greyc!30, line width=0.3pt},
  axis line style={dark!70, line width=0.5pt},
  every axis plot/.append style={line width=0.9pt, mark size=1.4pt},
  legend style={font=\footnotesize, draw=greyc!70, fill=white,
                at={(0.5,-0.50)}, anchor=north, legend columns=7,
                /tikz/every even column/.append style={column sep=6pt}},
]
\nextgroupplot[xmin=1.3, xmax=25.7, xtick={2,6,10,14,18,22}]
\addplot[specc, mark=*, mark options={fill=specc}] coordinates {(2,96.2) (3,95.1) (4,95.1) (5,95.0) (6,95.0) (7,94.5) (8,94.6) (9,94.4) (10,94.5) (11,93.8) (12,94.0) (13,93.4) (14,93.5) (15,92.2) (16,92.4) (17,92.4) (18,92.4) (19,92.0) (20,92.4) (21,91.5) (22,92.9) (24,89.9) (25,66.7)};
\addplot[Bblue, mark=square*, mark options={fill=Bblue!35}] coordinates {(2,94.9) (3,93.4) (4,93.3) (5,92.9) (6,92.9) (7,92.2) (8,92.2) (9,90.8) (10,90.7) (11,88.9) (12,88.9) (13,85.8) (14,85.7) (15,80.4) (16,80.4) (17,71.4) (18,71.2) (19,60.3) (20,60.1) (21,52.8) (22,53.3) (24,37.7) (25,0.0)};
\addplot[Ared, mark=triangle*, mark options={fill=Ared!35}] coordinates {(2,92.2) (3,89.9) (4,89.8) (5,89.2) (6,89.2) (7,88.2) (8,88.2) (9,86.0) (10,85.9) (11,83.2) (12,83.1) (13,78.2) (14,78.2) (15,70.1) (16,69.9) (17,57.2) (18,56.9) (19,42.1) (20,41.8) (21,26.5) (22,25.9) (24,16.4) (25,16.7)};
\addplot[mixc, mark=diamond*, mark options={fill=mixc!35}] coordinates {(2,82.2) (3,77.1) (4,76.8) (5,75.5) (6,75.4) (7,73.1) (8,73.1) (9,68.1) (10,68.0) (11,62.4) (12,62.3) (13,54.0) (14,53.9) (15,48.4) (16,48.0) (17,40.5) (18,40.2) (19,34.1) (20,33.8) (21,23.9) (22,22.6) (24,15.1) (25,0.0)};
\addplot[Ared!55, mark=pentagon*, mark options={fill=Ared!25}] coordinates {(2,61.2) (3,50.3) (4,49.5) (5,46.8) (6,46.7) (7,42.0) (8,41.8) (9,31.2) (10,31.0) (11,19.4) (12,19.3) (13,7.4) (14,7.4) (15,3.4) (16,3.4) (17,1.1) (18,1.0) (19,0.6) (20,0.6) (21,0.3) (22,0.3) (24,0.0) (25,0.0)};
\addplot[Bblue!55, mark=x, mark size=2.0pt] coordinates {(2,34.6) (3,21.5) (4,21.8) (5,19.3) (6,19.4) (7,15.7) (8,15.7) (9,12.9) (10,12.9) (11,8.3) (12,8.4) (13,6.5) (14,6.5) (15,2.7) (16,2.7) (17,1.2) (18,1.2) (19,0.5) (20,0.5) (21,0.3) (22,0.3) (24,0.0) (25,0.0)};
\addplot[greyc, mark=+, mark size=2.0pt] coordinates {(2,0.0) (3,0.0) (4,0.0) (5,0.0) (6,0.0) (7,0.0) (8,0.0) (9,0.0) (10,0.0) (11,0.0) (12,0.0) (13,0.0) (14,0.0) (15,0.0) (16,0.0) (17,0.0) (18,0.0) (19,0.0) (20,0.0) (21,0.0) (22,0.0) (24,0.0) (25,0.0)};
\node[font=\small, anchor=north] at (rel axis cs:0.5,-0.33) {(a) All codes};

\nextgroupplot[xmin=1.4, xmax=14.6, xtick={2,4,6,8,10,12,14}]
\addplot[specc, mark=*, mark options={fill=specc}] coordinates {(2,98.9) (3,97.6) (4,97.6) (5,97.6) (6,97.6) (7,96.4) (8,96.9) (9,96.4) (10,96.2) (11,92.5) (12,97.3) (14,100.0)};
\addlegendentry{BB-Net (ours)}
\addplot[Bblue, mark=square*, mark options={fill=Bblue!35}] coordinates {(2,100.0) (3,100.0) (4,100.0) (5,100.0) (6,100.0) (7,100.0) (8,100.0) (9,100.0) (10,100.0) (11,100.0) (12,100.0) (14,100.0)};
\addlegendentry{QDistEvol}
\addplot[Ared, mark=triangle*, mark options={fill=Ared!35}] coordinates {(2,100.0) (3,100.0) (4,100.0) (5,100.0) (6,100.0) (7,100.0) (8,100.0) (9,100.0) (10,100.0) (11,100.0) (12,100.0) (14,100.0)};
\addlegendentry{QDistRnd}
\addplot[mixc, mark=diamond*, mark options={fill=mixc!35}] coordinates {(2,100.0) (3,99.9) (4,99.9) (5,99.9) (6,99.9) (7,99.8) (8,99.8) (9,99.4) (10,99.4) (11,100.0) (12,100.0) (14,100.0)};
\addlegendentry{BP-OSD}
\addplot[Ared!55, mark=pentagon*, mark options={fill=Ared!25}] coordinates {(2,98.4) (3,97.5) (4,97.4) (5,97.1) (6,97.0) (7,95.9) (8,96.1) (9,89.8) (10,90.6) (11,75.0) (12,78.4) (14,100.0)};
\addlegendentry{UEStim}
\addplot[Bblue!55, mark=x, mark size=2.0pt] coordinates {(2,64.0) (3,44.1) (4,47.1) (5,40.7) (6,41.4) (7,27.5) (8,27.2) (9,30.7) (10,32.1) (11,22.5) (12,24.3) (14,0.0)};
\addlegendentry{CCStim}
\addplot[greyc, mark=+, mark size=2.0pt] coordinates {(2,0.0) (3,0.1) (4,0.1) (5,0.1) (6,0.1) (7,0.0) (8,0.0) (9,0.0) (10,0.0) (11,0.0) (12,0.0) (14,0.0)};
\addlegendentry{GEStim}
\node[font=\small, anchor=north] at (rel axis cs:0.5,-0.33) {(b) Certified distance};

\nextgroupplot[xmin=4.3, xmax=25.7, xtick={5,9,13,17,21,25}]
\addplot[specc, mark=*, mark options={fill=specc}] coordinates {(5,94.4) (6,94.4) (7,94.3) (8,94.3) (9,94.3) (10,94.4) (11,93.9) (12,94.0) (13,93.4) (14,93.5) (15,92.2) (16,92.4) (17,92.4) (18,92.4) (19,92.0) (20,92.4) (21,91.5) (22,92.9) (24,89.9) (25,66.7)};
\addplot[Bblue, mark=square*, mark options={fill=Bblue!35}] coordinates {(5,91.4) (6,91.4) (7,91.2) (8,91.1) (9,90.3) (10,90.2) (11,88.8) (12,88.7) (13,85.8) (14,85.7) (15,80.4) (16,80.4) (17,71.4) (18,71.2) (19,60.3) (20,60.1) (21,52.8) (22,53.3) (24,37.7) (25,0.0)};
\addplot[Ared, mark=triangle*, mark options={fill=Ared!35}] coordinates {(5,86.9) (6,86.9) (7,86.6) (8,86.6) (9,85.2) (10,85.2) (11,83.0) (12,82.9) (13,78.2) (14,78.2) (15,70.1) (16,69.9) (17,57.2) (18,56.9) (19,42.1) (20,41.8) (21,26.5) (22,25.9) (24,16.4) (25,16.7)};
\addplot[mixc, mark=diamond*, mark options={fill=mixc!35}] coordinates {(5,70.3) (6,70.3) (7,69.5) (8,69.5) (9,66.4) (10,66.3) (11,61.9) (12,61.8) (13,54.0) (14,53.9) (15,48.4) (16,48.0) (17,40.5) (18,40.2) (19,34.1) (20,33.8) (21,23.9) (22,22.6) (24,15.1) (25,0.0)};
\addplot[Ared!55, mark=pentagon*, mark options={fill=Ared!25}] coordinates {(5,36.1) (6,36.1) (7,34.6) (8,34.6) (9,28.0) (10,27.9) (11,18.5) (12,18.5) (13,7.3) (14,7.3) (15,3.4) (16,3.4) (17,1.1) (18,1.0) (19,0.6) (20,0.6) (21,0.3) (22,0.3) (24,0.0) (25,0.0)};
\addplot[Bblue!55, mark=x, mark size=2.0pt] coordinates {(5,14.7) (6,14.7) (7,14.1) (8,14.2) (9,11.9) (10,11.9) (11,8.1) (12,8.1) (13,6.5) (14,6.5) (15,2.7) (16,2.7) (17,1.2) (18,1.2) (19,0.5) (20,0.5) (21,0.3) (22,0.3) (24,0.0) (25,0.0)};
\addplot[greyc, mark=+, mark size=2.0pt] coordinates {(5,0.0) (6,0.0) (7,0.0) (8,0.0) (9,0.0) (10,0.0) (11,0.0) (12,0.0) (13,0.0) (14,0.0) (15,0.0) (16,0.0) (17,0.0) (18,0.0) (19,0.0) (20,0.0) (21,0.0) (22,0.0) (24,0.0) (25,0.0)};
\node[font=\small, anchor=north] at (rel axis cs:0.5,-0.33) {(c) Upper bound distance};
\end{groupplot}
\end{tikzpicture}%
}
\vspace{-12pt}
\caption{The \(x\)-axis denotes the minimum distance threshold \(t\), and the \(y\)-axis reports the accuracy computed over codes whose distance satisfies \(d \ge t\). Thus, moving to the right restricts the evaluation to increasingly high-distance codes. This was examined on three subsets: the whole benchmark, certified exact distances, and upper bound, unresolved distances.}
\label{fig:bbnet_acc}
\end{figure}

\vspace{-5pt}
Of the 5,870 test codes, the MILP-based exact solver certified the code distance for 2,366 instances. The remaining instances were substantially more difficult to certify and generally exhibited larger heuristic upper bounds. None of the certified codes had a target distance greater than 14. We consequently report results separately for exact codes and for the upper bounds codes. On the certified subset (\autoref{tab:benchmark_certified}), BB-Net's accuracy is 98.94\% compared to 100\% for QDistEvol, QDistRnd, and BP-OSD. UEStim with 98.44\%, CCStim with 64.03\% and GEStim with 0.04\%. On the unresolved upper bound subset (\autoref{tab:benchmark_unresolved}), BB-Net achieves an accuracy of 94.66\% and an RMSE of 1.380, compared to 91.38\% accuracy and an RMSE of 1.923 for the strongest heuristic baseline, QDistEvol. These results indicate that BB-Net is particularly effective on the high-targets upper bound distance instances for which both exact solvers and heuristics struggle. Accuracy across distances for the different subsets is presented in \autoref{fig:bbnet_acc}. Selected challenging codes are presented in \autoref{tab:bbnet_challenge}.

%
%
\vspace{-5pt}

\begin{table}[!h]
\centering
\caption{Certified distance codes, $d\leq14$ (2,366 codes)}
\label{tab:benchmark_certified}
\small
\setlength{\tabcolsep}{2pt}
\begin{tabular}{l rrrrrr | r}
\toprule
& QDistRnd & QDistEvol & BP-OSD & GEStim & CCStim & UEStim & BB-Net (ours) \\
\midrule
Trials Per Code                    & 200 & 200 & 200 & 1 & 1 & 3 & 1  \\
Time Per Code [s] ($\downarrow$)   & 5.0 & 1.6 & 3.8 & 1.0 & 1.0 & 2.1 & \textbf{0.005} \\
Total Time [s] ($\downarrow$)      & 11,887 & 3,883 & 9,032 & 2,406 & 2,407 & 4,898 & \textbf{10.8} \\
RMSE ($\downarrow$)                & \textbf{0.000} & \textbf{0.000} & 0.699 & 11.453 & 7.434 & 2.641 & 0.594 \\
Accuracy ($\uparrow$)              & \textbf{100.00\%} & \textbf{100.00\%} & 99.96\% & 0.04\% & 64.03\% & 98.44\% & 98.94\% \\
\bottomrule
\end{tabular}
\end{table}

\vspace{-5pt}
\begin{table}[!h]
\centering
\caption{Upper bound distance codes (3,504 codes)}
\label{tab:benchmark_unresolved}
\small
\setlength{\tabcolsep}{2pt}
\begin{tabular}{l rrrrrr | r}
\toprule
& QDistRnd & QDistEvol & BP-OSD & GEStim & CCStim & UEStim & BB-Net (ours) \\
\midrule
Trials Per Code                    & 200 & 200 & 200 & 1 & 1 & 3 & 1  \\
Time Per Code [s] ($\downarrow$)   & 2.1 & 0.6 & 29.5 & 1.3 & 1.4 & 20.5 & \textbf{0.005} \\
Total Time [s] ($\downarrow$)      & 7,496 & 2,175 & 103,243 & 4,535 & 4,758 & 71,877 & \textbf{16.0} \\
RMSE ($\downarrow$)                & 3.002 & 1.923 & 30.181 & 42.355 & 35.362 & 37.190 & \textbf{1.380} \\
Accuracy ($\uparrow$)              & 86.93\% & 91.38\% & 70.26\% & 0.00\% & 14.73\% & 36.10\% & \textbf{94.66\%} \\
\bottomrule
\end{tabular}
\end{table}

\newpage
\subsection{Distance Screening}
\label{sec:distance_screening}
\vspace{-5pt}
A common and crucial strategy in code-searching pipelines is distance screening: limited computational budget is first used to filter candidates above a certain potential distance, after which computationally intensive heuristics or exact solvers are applied only to the most promising codes to tighten bounds or obtain certificates.
To evaluate BB-Net's suitability for distance screening, we compute ROC-AUC at distance thresholds $t$, using $P(d \ge t)$ as the ranking score and measuring how well codes above and below each threshold are separated (\autoref{tab:auc_threshold}). BB-Net achieves slightly below QDistEvol and QDistRnd at intermediate thresholds and comparable at the highest ones, while requiring only a fraction of their inference time (reported in \autoref{tab:benchmark_200}).

\vspace{-5pt}
\begin{table}[!h]
\centering
\caption{AUC for distance screening over the test benchmark. Best value in each column in bold, second best
underlined (higher is better).}
\label{tab:auc_threshold}
\small
\setlength{\tabcolsep}{4.5pt}
\begin{tabular}{l cccccccccc}
\toprule
 & $d\!\ge\!4$ & $d\!\ge\!8$ & $d\!\ge\!10$ & $d\!\ge\!12$ & $d\!\ge\!14$
       & $d\!\ge\!16$ & $d\!\ge\!18$ & $d\!\ge\!20$ & $d\!\ge\!22$ & $d\!\ge\!24$ \\
\midrule
BB-Net    & \textbf{1.000} & \underline{0.998} & 0.996 & 0.993 & 0.989
          & \underline{0.989} & 0.988 & 0.987 & \underline{0.990} & \underline{0.991} \\
QDistEvol & \textbf{1.000} & \textbf{1.000} & \textbf{1.000} & \textbf{1.000}
          & \underline{0.999} & \textbf{0.998} & \textbf{0.998} & \textbf{0.995}
          & \textbf{0.991} & \textbf{0.992} \\
QDistRnd  & \textbf{1.000} & \textbf{1.000} & \textbf{1.000} & \textbf{1.000}
          & \textbf{1.000} & \textbf{0.998} & \underline{0.997} & \underline{0.993}
          & \textbf{0.991} & 0.990 \\
BP-OSD    & \textbf{1.000} & \textbf{1.000} & \textbf{1.000} & \underline{0.997}
          & 0.987 & 0.967 & 0.950 & 0.936 & 0.931 & 0.926 \\
UEStim    & 0.996 & 0.996 & \underline{0.997} & 0.992 & 0.974 & 0.944 & 0.917
          & 0.900 & 0.890 & 0.883 \\
GEStim    & \underline{0.998} & 0.946 & 0.913 & 0.892 & 0.876 & 0.871 & 0.855
          & 0.849 & 0.845 & 0.845 \\
CCStim    & 0.991 & 0.973 & 0.934 & 0.924 & 0.908 & 0.909 & 0.895 & 0.889
          & 0.884 & 0.893 \\
\bottomrule
\end{tabular}
\end{table}

\subsection{Ablation}
\label{sec:ablation}
\begin{figure}[!h]
\centering
\definecolor{Ared} {RGB}{198, 60, 52}%
\definecolor{Bblue}{RGB}{ 52, 92,168}%
\definecolor{specc}{RGB}{ 20,106, 96}%
\definecolor{dark} {RGB}{ 60, 60, 60}%
\definecolor{greyc}{RGB}{150,150,150}%
\resizebox{0.8\textwidth}{!}{%
\begin{tikzpicture}
\begin{groupplot}[
  group style={group size=3 by 1, horizontal sep=0.55cm,
               y descriptions at=edge left},
  width=6.1cm, height=5.3cm,
  xmin=3, xmax=92, xtick={10,30,50,70,90},
  ymin=0.60, ymax=2.75, ytick={0.8,1.2,1.6,2.0,2.4},
  ylabel={NLL ($\downarrow$)},
  xlabel style={font=\small, align=center}, ylabel style={font=\small},
  tick label style={font=\footnotesize},
  grid=major, grid style={greyc!30, line width=0.3pt},
  axis line style={dark!70, line width=0.5pt},
  every axis plot/.append style={line width=0.9pt, mark size=1.4pt},
  legend style={font=\footnotesize, draw=greyc!70, fill=white,
                at={(1.59,-0.42)}, anchor=north, legend columns=4,
                /tikz/every even column/.append style={column sep=6pt}},
]
\nextgroupplot[xlabel={Epoch\\[2pt](a) Augmentation}]
\addplot[Bblue, mark=square*, mark options={fill=Bblue!35}, mark repeat=12] coordinates {(5,1.8232) (6,1.8505) (7,1.7232) (8,1.7008) (9,1.6526) (10,1.6113) (11,1.5637) (12,1.5512) (13,1.5289) (14,1.4821) (15,1.4759) (16,1.4371) (17,1.4066) (18,1.3945) (19,1.3866) (20,1.3598) (21,1.3459) (22,1.3458) (23,1.3218) (24,1.3153) (25,1.3208) (26,1.3090) (27,1.3018) (28,1.2846) (29,1.2921) (30,1.2880) (31,1.2741) (32,1.2581) (33,1.2657) (34,1.2635) (35,1.2528) (36,1.2377) (37,1.2395) (38,1.2318) (39,1.2260) (40,1.2255) (41,1.2192) (42,1.2110) (43,1.2244) (44,1.2066) (45,1.2046) (46,1.2094) (47,1.2063) (48,1.1888) (49,1.1924) (50,1.1894) (51,1.1919) (52,1.1871) (53,1.1857) (54,1.1770) (55,1.1791) (56,1.1729) (57,1.1748) (58,1.1798) (59,1.1712) (60,1.1670) (61,1.1739) (62,1.1668) (63,1.1682) (64,1.1677) (65,1.1661) (66,1.1665) (67,1.1637) (68,1.1622) (69,1.1618) (70,1.1615) (71,1.1690) (72,1.1646) (73,1.1629) (74,1.1590) (75,1.1611) (76,1.1603) (77,1.1644) (78,1.1647) (79,1.1639) (80,1.1636) (81,1.1633) (82,1.1636) (83,1.1632) (84,1.1629) (85,1.1634) (86,1.1635) (87,1.1636) (88,1.1634) (89,1.1636) (90,1.1636)};
\addlegendentry{BB-Net}
\addplot[Bblue!45, mark=square, mark options={fill=white}, mark repeat=12] coordinates {(5,1.7870) (6,1.7081) (7,1.7016) (8,1.5673) (9,1.5291) (10,1.4884) (11,1.4739) (12,1.5016) (13,1.4422) (14,1.4455) (15,1.4134) (16,1.4492) (17,1.4142) (18,1.4628) (19,1.4441) (20,1.4907) (21,1.5151) (22,1.5580) (23,1.6008) (24,1.6349) (25,1.6754) (26,1.7595) (27,1.8252) (28,1.8655) (29,1.9779) (30,2.0519) (31,2.1131) (32,2.1324) (33,2.2434) (34,2.3465) (35,2.3694) (36,2.4252) (37,2.5188) (38,2.5767) (39,2.5966) (40,2.7321) (41,2.7316) (42,2.7939) (43,2.8188) (44,2.8587) (45,2.9160) (46,2.9402) (47,3.0325) (48,3.0353) (49,3.1592) (50,3.1909) (51,3.1795) (52,3.2460) (53,3.2653) (54,3.3025) (55,3.3853) (56,3.4291) (57,3.4662) (58,3.4757) (59,3.5714) (60,3.6446) (61,3.6781) (62,3.7220) (63,3.7271) (64,3.7873) (65,3.8525) (66,3.8570) (67,3.9066) (68,3.9806) (69,4.0224) (70,4.0777) (71,4.0894) (72,4.1810) (73,4.1887) (74,4.2513) (75,4.2857) (76,4.3181) (77,4.3638) (78,4.4012) (79,4.4382) (80,4.4437) (81,4.4784) (82,4.4897) (83,4.4987) (84,4.5092) (85,4.5162) (86,4.5225) (87,4.5246) (88,4.5253) (89,4.5264) (90,4.5263)};
\addlegendentry{BB-Net (ablated)}
\addplot[specc, mark=*, mark options={fill=specc}, mark repeat=12] coordinates {(5,1.6372) (6,1.5353) (7,1.3364) (8,1.3686) (9,1.2393) (10,1.1656) (11,1.1163) (12,1.1204) (13,1.1248) (14,0.9843) (15,0.9259) (16,1.0161) (17,0.9899) (18,0.9316) (19,0.8848) (20,0.9121) (21,0.8733) (22,0.8773) (23,0.8315) (24,0.9181) (25,0.8653) (26,0.8845) (27,0.8364) (28,0.9057) (29,0.7686) (30,0.7732) (31,0.8205) (32,0.7737) (33,0.8104) (34,0.7473) (35,0.7443) (36,0.7445) (37,0.7338) (38,0.7517) (39,0.7095) (40,0.8116) (41,0.7721) (42,0.7158) (43,0.7847) (44,0.7493) (45,0.8356) (46,0.7115) (47,0.7086) (48,0.7245) (49,0.8168) (50,0.7183) (51,0.7605) (52,0.7820) (53,0.7561) (54,0.8048) (55,0.7729) (56,0.7902) (57,0.8018) (58,0.8186) (59,0.8303) (60,0.8733) (61,0.8644) (62,0.8510) (63,0.8924) (64,0.9186) (65,0.8976) (66,0.9001) (67,0.9392) (68,0.9654) (69,0.9621) (70,0.9848) (71,1.0193) (72,1.0288) (73,1.0440) (74,1.0383) (75,1.0456) (76,1.0748) (77,1.0753) (78,1.0930) (79,1.1153) (80,1.1045) (81,1.1134) (82,1.1142) (83,1.1288) (84,1.1319) (85,1.1334) (86,1.1404) (87,1.1420) (88,1.1411) (89,1.1421) (90,1.1418)};
\addlegendentry{BB-Net-Tanner}
\addplot[specc!45, mark=o, mark options={fill=white}, mark repeat=12] coordinates {(5,1.5813) (6,1.4079) (7,1.3139) (8,1.2247) (9,1.2280) (10,1.1332) (11,1.1152) (12,1.0919) (13,1.0743) (14,0.9920) (15,0.9850) (16,0.9519) (17,1.0159) (18,0.9667) (19,0.9563) (20,1.0577) (21,1.0559) (22,1.0706) (23,1.0045) (24,1.0687) (25,1.1183) (26,1.1492) (27,1.3035) (28,1.2438) (29,1.2766) (30,1.2710) (31,1.4796) (32,1.4235) (33,1.4534) (34,1.5249) (35,1.5125) (36,1.5529) (37,1.5272) (38,1.6629) (39,1.6394) (40,1.7446) (41,1.6580) (42,1.6480) (43,1.7867) (44,1.8797) (45,1.8320) (46,1.8083) (47,1.7853) (48,1.9135) (49,1.8964) (50,1.9651) (51,1.9744) (52,1.9962) (53,1.9814) (54,2.0020) (55,2.0601) (56,2.0174) (57,2.0925) (58,2.2160) (59,2.1427) (60,2.1205) (61,2.1583) (62,2.1906) (63,2.2256) (64,2.2782) (65,2.2663) (66,2.3085) (67,2.2902) (68,2.3115) (69,2.3232) (70,2.3143) (71,2.3419) (72,2.3521) (73,2.3687) (74,2.3771) (75,2.3651) (76,2.3888) (77,2.4186) (78,2.3986) (79,2.3919) (80,2.3889) (81,2.3881) (82,2.3801) (83,2.3841) (84,2.3835) (85,2.3671) (86,2.3824) (87,2.3856) (88,2.3904) (89,2.3867) (90,2.3868)};
\addlegendentry{BB-Net-Tanner (ablated)}
\nextgroupplot[xlabel={Epoch\\[2pt](b) Shift canonicalisation}]
\addplot[Bblue, mark=square*, mark options={fill=Bblue!35}, mark repeat=12, forget plot] coordinates {(5,1.8232) (6,1.8505) (7,1.7232) (8,1.7008) (9,1.6526) (10,1.6113) (11,1.5637) (12,1.5512) (13,1.5289) (14,1.4821) (15,1.4759) (16,1.4371) (17,1.4066) (18,1.3945) (19,1.3866) (20,1.3598) (21,1.3459) (22,1.3458) (23,1.3218) (24,1.3153) (25,1.3208) (26,1.3090) (27,1.3018) (28,1.2846) (29,1.2921) (30,1.2880) (31,1.2741) (32,1.2581) (33,1.2657) (34,1.2635) (35,1.2528) (36,1.2377) (37,1.2395) (38,1.2318) (39,1.2260) (40,1.2255) (41,1.2192) (42,1.2110) (43,1.2244) (44,1.2066) (45,1.2046) (46,1.2094) (47,1.2063) (48,1.1888) (49,1.1924) (50,1.1894) (51,1.1919) (52,1.1871) (53,1.1857) (54,1.1770) (55,1.1791) (56,1.1729) (57,1.1748) (58,1.1798) (59,1.1712) (60,1.1670) (61,1.1739) (62,1.1668) (63,1.1682) (64,1.1677) (65,1.1661) (66,1.1665) (67,1.1637) (68,1.1622) (69,1.1618) (70,1.1615) (71,1.1690) (72,1.1646) (73,1.1629) (74,1.1590) (75,1.1611) (76,1.1603) (77,1.1644) (78,1.1647) (79,1.1639) (80,1.1636) (81,1.1633) (82,1.1636) (83,1.1632) (84,1.1629) (85,1.1634) (86,1.1635) (87,1.1636) (88,1.1634) (89,1.1636) (90,1.1636)};
\addplot[Bblue!45, mark=square, mark options={fill=white}, mark repeat=12, forget plot] coordinates {(5,1.9899) (6,2.0851) (7,1.9463) (8,1.9213) (9,2.0474) (10,1.8282) (11,1.8364) (12,1.8377) (13,1.7704) (14,1.8023) (15,1.7258) (16,1.7025) (17,1.7352) (18,1.7282) (19,1.6881) (20,1.6787) (21,1.6611) (22,1.6574) (23,1.6347) (24,1.6370) (25,1.6235) (26,1.6295) (27,1.6444) (28,1.6013) (29,1.5923) (30,1.5981) (31,1.5917) (32,1.5835) (33,1.5735) (34,1.5672) (35,1.5615) (36,1.5674) (37,1.5628) (38,1.5432) (39,1.5471) (40,1.5396) (41,1.5261) (42,1.5249) (43,1.5240) (44,1.5175) (45,1.5165) (46,1.5217) (47,1.5117) (48,1.5161) (49,1.5058) (50,1.5064) (51,1.5048) (52,1.4988) (53,1.4935) (54,1.4913) (55,1.4997) (56,1.4875) (57,1.4797) (58,1.4943) (59,1.4856) (60,1.4789) (61,1.4746) (62,1.4696) (63,1.4682) (64,1.4634) (65,1.4639) (66,1.4639) (67,1.4692) (68,1.4566) (69,1.4544) (70,1.4561) (71,1.4596) (72,1.4503) (73,1.4486) (74,1.4487) (75,1.4444) (76,1.4448) (77,1.4414) (78,1.4408) (79,1.4404) (80,1.4412) (81,1.4396) (82,1.4381) (83,1.4366) (84,1.4373) (85,1.4369) (86,1.4376) (87,1.4372) (88,1.4371) (89,1.4371) (90,1.4370)};
\addplot[specc, mark=*, mark options={fill=specc}, mark repeat=12, forget plot] coordinates {(5,1.6372) (6,1.5353) (7,1.3364) (8,1.3686) (9,1.2393) (10,1.1656) (11,1.1163) (12,1.1204) (13,1.1248) (14,0.9843) (15,0.9259) (16,1.0161) (17,0.9899) (18,0.9316) (19,0.8848) (20,0.9121) (21,0.8733) (22,0.8773) (23,0.8315) (24,0.9181) (25,0.8653) (26,0.8845) (27,0.8364) (28,0.9057) (29,0.7686) (30,0.7732) (31,0.8205) (32,0.7737) (33,0.8104) (34,0.7473) (35,0.7443) (36,0.7445) (37,0.7338) (38,0.7517) (39,0.7095) (40,0.8116) (41,0.7721) (42,0.7158) (43,0.7847) (44,0.7493) (45,0.8356) (46,0.7115) (47,0.7086) (48,0.7245) (49,0.8168) (50,0.7183) (51,0.7605) (52,0.7820) (53,0.7561) (54,0.8048) (55,0.7729) (56,0.7902) (57,0.8018) (58,0.8186) (59,0.8303) (60,0.8733) (61,0.8644) (62,0.8510) (63,0.8924) (64,0.9186) (65,0.8976) (66,0.9001) (67,0.9392) (68,0.9654) (69,0.9621) (70,0.9848) (71,1.0193) (72,1.0288) (73,1.0440) (74,1.0383) (75,1.0456) (76,1.0748) (77,1.0753) (78,1.0930) (79,1.1153) (80,1.1045) (81,1.1134) (82,1.1142) (83,1.1288) (84,1.1319) (85,1.1334) (86,1.1404) (87,1.1420) (88,1.1411) (89,1.1421) (90,1.1418)};
\addplot[specc!45, mark=o, mark options={fill=white}, mark repeat=12, forget plot] coordinates {(5,1.7189) (6,1.5514) (7,1.4461) (8,1.3714) (9,1.2549) (10,1.3198) (11,1.2948) (12,1.2067) (13,1.1864) (14,1.0678) (15,1.1009) (16,1.0184) (17,0.9685) (18,0.9568) (19,1.0832) (20,0.9566) (21,0.9665) (22,0.9767) (23,0.9028) (24,0.8879) (25,0.9214) (26,0.9366) (27,0.9490) (28,0.8326) (29,0.8459) (30,0.9111) (31,0.9391) (32,0.8527) (33,0.8588) (34,0.8822) (35,0.8466) (36,0.7897) (37,0.8107) (38,0.8472) (39,0.7880) (40,0.9014) (41,0.8089) (42,0.8030) (43,0.8750) (44,0.8004) (45,0.8309) (46,0.8062) (47,0.8048) (48,0.7878) (49,0.8592) (50,0.8538) (51,0.8307) (52,0.8120) (53,0.8547) (54,0.8334) (55,0.8340) (56,0.9255) (57,0.8812) (58,0.9463) (59,0.9256) (60,0.9747) (61,0.9697) (62,0.9394) (63,0.9923) (64,1.0079) (65,1.0160) (66,0.9879) (67,1.0331) (68,1.0358) (69,1.0474) (70,1.0486) (71,1.0853) (72,1.0796) (73,1.1088) (74,1.1350) (75,1.1359) (76,1.1399) (77,1.1654) (78,1.1754) (79,1.1668) (80,1.1790) (81,1.1923) (82,1.1877) (83,1.1991) (84,1.2121) (85,1.2154) (86,1.2190) (87,1.2170) (88,1.2178) (89,1.2161) (90,1.2159)};
\nextgroupplot[xlabel={Epoch\\[2pt](c) FiLM conditioning}]
\addplot[Bblue, mark=square*, mark options={fill=Bblue!35}, mark repeat=12, forget plot] coordinates {(5,1.8232) (6,1.8505) (7,1.7232) (8,1.7008) (9,1.6526) (10,1.6113) (11,1.5637) (12,1.5512) (13,1.5289) (14,1.4821) (15,1.4759) (16,1.4371) (17,1.4066) (18,1.3945) (19,1.3866) (20,1.3598) (21,1.3459) (22,1.3458) (23,1.3218) (24,1.3153) (25,1.3208) (26,1.3090) (27,1.3018) (28,1.2846) (29,1.2921) (30,1.2880) (31,1.2741) (32,1.2581) (33,1.2657) (34,1.2635) (35,1.2528) (36,1.2377) (37,1.2395) (38,1.2318) (39,1.2260) (40,1.2255) (41,1.2192) (42,1.2110) (43,1.2244) (44,1.2066) (45,1.2046) (46,1.2094) (47,1.2063) (48,1.1888) (49,1.1924) (50,1.1894) (51,1.1919) (52,1.1871) (53,1.1857) (54,1.1770) (55,1.1791) (56,1.1729) (57,1.1748) (58,1.1798) (59,1.1712) (60,1.1670) (61,1.1739) (62,1.1668) (63,1.1682) (64,1.1677) (65,1.1661) (66,1.1665) (67,1.1637) (68,1.1622) (69,1.1618) (70,1.1615) (71,1.1690) (72,1.1646) (73,1.1629) (74,1.1590) (75,1.1611) (76,1.1603) (77,1.1644) (78,1.1647) (79,1.1639) (80,1.1636) (81,1.1633) (82,1.1636) (83,1.1632) (84,1.1629) (85,1.1634) (86,1.1635) (87,1.1636) (88,1.1634) (89,1.1636) (90,1.1636)};
\addplot[Bblue!45, mark=square, mark options={fill=white}, mark repeat=12, forget plot] coordinates {(5,1.9768) (6,1.9202) (7,1.8822) (8,1.8681) (9,1.8519) (10,1.7922) (11,1.7815) (12,1.7225) (13,1.7003) (14,1.6803) (15,1.6627) (16,1.6253) (17,1.6158) (18,1.5867) (19,1.5721) (20,1.5428) (21,1.5345) (22,1.5095) (23,1.4918) (24,1.4778) (25,1.4711) (26,1.4466) (27,1.4391) (28,1.4154) (29,1.4070) (30,1.4019) (31,1.3861) (32,1.3855) (33,1.3669) (34,1.3644) (35,1.3534) (36,1.3403) (37,1.3377) (38,1.3257) (39,1.3279) (40,1.3197) (41,1.3097) (42,1.3030) (43,1.2951) (44,1.2946) (45,1.2882) (46,1.2806) (47,1.2767) (48,1.2758) (49,1.2706) (50,1.2677) (51,1.2563) (52,1.2608) (53,1.2532) (54,1.2520) (55,1.2517) (56,1.2447) (57,1.2422) (58,1.2373) (59,1.2425) (60,1.2343) (61,1.2311) (62,1.2335) (63,1.2299) (64,1.2298) (65,1.2276) (66,1.2288) (67,1.2230) (68,1.2254) (69,1.2248) (70,1.2217) (71,1.2231) (72,1.2237) (73,1.2217) (74,1.2200) (75,1.2197) (76,1.2196) (77,1.2202) (78,1.2180) (79,1.2184) (80,1.2190) (81,1.2183) (82,1.2181) (83,1.2182) (84,1.2184) (85,1.2187) (86,1.2182) (87,1.2183) (88,1.2183) (89,1.2183) (90,1.2184)};
\addplot[specc, mark=*, mark options={fill=specc}, mark repeat=12, forget plot] coordinates {(5,1.6372) (6,1.5353) (7,1.3364) (8,1.3686) (9,1.2393) (10,1.1656) (11,1.1163) (12,1.1204) (13,1.1248) (14,0.9843) (15,0.9259) (16,1.0161) (17,0.9899) (18,0.9316) (19,0.8848) (20,0.9121) (21,0.8733) (22,0.8773) (23,0.8315) (24,0.9181) (25,0.8653) (26,0.8845) (27,0.8364) (28,0.9057) (29,0.7686) (30,0.7732) (31,0.8205) (32,0.7737) (33,0.8104) (34,0.7473) (35,0.7443) (36,0.7445) (37,0.7338) (38,0.7517) (39,0.7095) (40,0.8116) (41,0.7721) (42,0.7158) (43,0.7847) (44,0.7493) (45,0.8356) (46,0.7115) (47,0.7086) (48,0.7245) (49,0.8168) (50,0.7183) (51,0.7605) (52,0.7820) (53,0.7561) (54,0.8048) (55,0.7729) (56,0.7902) (57,0.8018) (58,0.8186) (59,0.8303) (60,0.8733) (61,0.8644) (62,0.8510) (63,0.8924) (64,0.9186) (65,0.8976) (66,0.9001) (67,0.9392) (68,0.9654) (69,0.9621) (70,0.9848) (71,1.0193) (72,1.0288) (73,1.0440) (74,1.0383) (75,1.0456) (76,1.0748) (77,1.0753) (78,1.0930) (79,1.1153) (80,1.1045) (81,1.1134) (82,1.1142) (83,1.1288) (84,1.1319) (85,1.1334) (86,1.1404) (87,1.1420) (88,1.1411) (89,1.1421) (90,1.1418)};
\addplot[specc!45, mark=o, mark options={fill=white}, mark repeat=12, forget plot] coordinates {(5,1.6436) (6,1.4117) (7,1.3818) (8,1.3528) (9,1.2913) (10,1.1675) (11,1.1976) (12,1.2712) (13,1.4161) (14,1.1153) (15,1.2265) (16,0.9859) (17,1.0638) (18,1.1069) (19,0.9205) (20,1.0679) (21,0.9198) (22,0.8800) (23,0.8964) (24,0.9391) (25,0.8537) (26,0.9617) (27,0.9167) (28,0.9384) (29,0.8720) (30,0.9260) (31,0.8750) (32,0.7572) (33,0.8130) (34,0.8314) (35,0.9473) (36,0.7930) (37,0.8228) (38,0.7609) (39,0.7409) (40,0.7748) (41,0.8013) (42,0.7890) (43,0.7428) (44,0.7766) (45,0.8687) (46,0.7854) (47,0.8010) (48,0.8221) (49,0.7495) (50,0.7927) (51,0.7809) (52,0.7690) (53,0.7456) (54,0.7600) (55,0.8116) (56,0.8414) (57,0.8196) (58,0.8309) (59,0.8041) (60,0.8344) (61,0.9108) (62,0.8751) (63,0.8473) (64,0.9223) (65,0.8706) (66,0.8988) (67,0.9630) (68,0.9560) (69,0.9824) (70,0.9693) (71,1.0267) (72,1.0097) (73,1.0234) (74,1.0157) (75,1.0600) (76,1.0705) (77,1.0984) (78,1.0888) (79,1.0884) (80,1.0979) (81,1.0858) (82,1.0917) (83,1.1138) (84,1.1115) (85,1.1226) (86,1.1226) (87,1.1224) (88,1.1216) (89,1.1217) (90,1.1229)};
\end{groupplot}
\end{tikzpicture}%
}
\vspace{-7pt}
\caption{Bold curves denote the best configuration, including augmentation, shift canonicalisation, and FiLM conditioning; lighter curves show the corresponding model with the component named in each panel removed.}
\label{fig:ablation}
\end{figure}
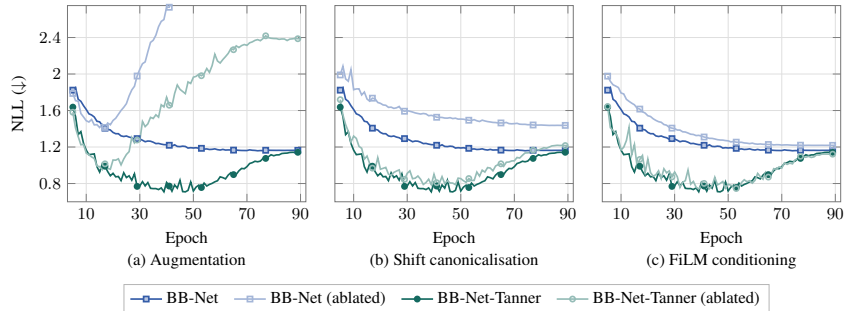

\vspace{-7pt}
To validate the proposed architectural design choices, we conduct a comparison of model variants and present them in \autoref{tab:bbnet_ablation} and \autoref{fig:ablation}. Specifically, we train them on the train set and evaluate their NLL on the validation set. We find that BB-Net with a tanner spectral branch, achieves NLL of 1.039 compare to 1.496 without it. Augmentation, shift-canonicalization and FiLM conditioning significantly contribute to the results, achieving NLL of 0.683. Unless stated otherwise, throughout the paper we use the term BB-Net to refer to the best-performing configuration.
%
\vspace{-10pt}

\vspace{-15pt}
\begin{table}[!h]
\centering
\small
\setlength{\tabcolsep}{5pt}
\renewcommand{\arraystretch}{1.05}
\caption{Ablation study of BB-Net on the validation set. Best-epoch NLL is reported.}
\label{tab:bbnet_ablation}
\begin{tabular}{ccc cc}
\toprule
\multicolumn{3}{c}{Component} & \multicolumn{2}{c}{NLL $\downarrow$} \\
\cmidrule(lr){1-3}\cmidrule(lr){4-5}
Aug. & Shift-Canon & FiLM & BB-Net & BB-Net-Tanner \\
\midrule
$\times$     & $\times$     & $\checkmark$ & $1.496 \pm 0.014$ & $1.039 \pm 0.019$ \\
$\checkmark$ & $\times$     & $\checkmark$ & $1.436 \pm 0.014$ & $0.762 \pm 0.012$ \\
$\times$     & $\checkmark$ & $\checkmark$ & $1.409 \pm 0.013$ & $0.924 \pm 0.026$ \\
$\checkmark$ & $\checkmark$ & $\times$     & $1.217 \pm 0.006$ & $0.712 \pm 0.025$ \\
\midrule
$\checkmark$ & $\checkmark$ & $\checkmark$ & $1.157 \pm 0.001$ & $\mathbf{0.683 \pm 0.010}$ \\
\bottomrule
\end{tabular}
\end{table}

\newpage
\section{Conclusion}
\vspace{-7pt}
In this work, we reframed the problem of quantum-code distance computation as an ML task. We introduced BB-Net, an attention-based architecture for predicting the distance of qLDPC codes from the BB family. BB-Net combines polynomial and Tanner-spectral representations of the sparse algebraic structure of BB codes and is trained on a dedicated dataset of upper bound distance-labeled instances by high-budget heuristics. Our test benchmark experiments show that BB-Net recovers the referenced distances more closely than the same heuristics with a budget of up to $10^4$ times more computational runtime, and that it succeeds on certified distances as well. These results establish learned distance prediction as a practical alternative to repeated computationally expensive distance estimation and enable substantially faster screening of candidate codes. By alleviating one of the primary computational bottlenecks in code search, BB-Net can accelerate the exploration and discovery of high-quality QEC codes, where large-scale searches currently rely heavily on costly heuristic distance evaluations.

\bibliography{iclr2027_conference}
\bibliographystyle{iclr2027_conference}

\appendix
\section{Experimental Results}
\label{app:experimental_results}

\subsection{Training and Inference Configuration}
All training experiments were conducted on an L40s GPU using PyTorch on the dataset we have created. Training hyperparametrs configuration are presented in \autoref{tab:bbnet_train_config}.
Both heuristics and BB-Net inference were conducted on a compute node equipped with two AMD EPYC 7763 64-Core processors, providing 128 physical CPU cores (256 hardware threads). 
Heuristics' invalid answers as defined in \autoref{sec:distance_finding} are also included in the results tables and affect their corresponding accuracy and RMSE.

In the ablation experiments, to match number of parameters when comparing BB-Net to BB-Net-Tanner, we set $L=16$ for BB-Net. Both models have approximately $13$M parameters.

\begin{table}[!h]
\centering
\caption{Hyperparameters used for training.}
\vspace{3pt}
\label{tab:bbnet_train_config}
\small

\begin{tabular}{l c }
\toprule
Params &
Values \\
\midrule

$L$ & 8 \\
$d_{\mathrm{model}}$ & 256 \\
lr & 3e-4 \\
Dropout & 0.1 \\
Epochs & 90 \\
Warmup Epochs & 9 \\
Batch Size & 256 \\
Optimizer & AdamW \\
Weight Decay & 0.01 \\

\bottomrule
\end{tabular}
\end{table}

\begin{table}[!h]
\centering
\caption{Selected test codes on which the heuristics overestimate the reference distance and BB-Net matches it. The reference is the minimum found by the same heuristics at 10,000 trials. GEStim, CCStim and UEStim failed to output a result.}
\label{tab:bbnet_challenge}
\scriptsize
\setlength{\tabcolsep}{1pt}
\renewcommand{\arraystretch}{1.18}

\resizebox{\textwidth}{!}{%
\begin{tabular}{l l l l l r r r r | r}
\toprule
$(N,\ell,m)$ &
$A(x,y)$ &
$B(x,y)$ &
$C(x,y)$ &
$D(x,y)$ &
GT &
BP-OSD &
QDistEvol &
QDistRnd &
BB-Net (ours) \\
\midrule

$(324,18,9)$ &
$x^{2}y^{2}+x^{9}y+x^{9}y^{8}+x^{13}y$ &
$x^{4}y^{8}+x^{8}y+x^{12}y^{4}+x^{16}y$ &
-- & -- &
18 & 44 & 45 & \textbf{18} & \textbf{18} \\

$(270,27,5)$ &
$y^{2}+x^{5}+x^{16}y+x^{26}y^{3}$ &
$x^{8}y^{3}+x^{10}y^{4}+x^{24}y^{3}+x^{26}y^{2}$ &
-- & -- &
25 & 35 & 38 & 37 & \textbf{25} \\

$(342,19,9)$ &
$x^{8}y^{2}+x^{16}y+x^{16}y^{3}$ &
$y^{5}+x^{2}y^{7}+x^{12}y^{3}$ &
$x^{12}y^{5}+x^{12}y^{7}$ &
$x^{6}y^{3}+x^{6}y^{5}$ &
24 & Fail & 36 & 26 & \textbf{24} \\

$(348,29,6)$ &
$x^{8}y^{2}+x^{13}y+x^{25}y^{3}$ &
$x^{2}y^{5}+x^{12}y^{4}+x^{16}y^{3}$ &
$x^{11}y^{4}+x^{22}y$ &
$x^{18}+x^{25}y^{3}$ &
24 & Fail & 36 & 44 & \textbf{24} \\

$(272,17,8)$ &
$xy^{2}+x^{4}y^{2}+x^{8}y^{3}+x^{9}y^{7}$ &
$y^{4}+x^{7}y^{2}+x^{8}y^{2}+x^{12}y^{2}$ &
-- & -- &
25 & 37 & 36 & 28 & \textbf{25} \\

$(360,30,6)$ &
$x^{2}+x^{7}+x^{19}y^{5}$ &
$x^{5}y^{2}+x^{12}y^{2}+x^{26}y$ &
-- & -- &
20 & 30 & 30 & 22 & \textbf{20} \\

$(360,18,10)$ &
$x^{7}y^{8}+x^{9}y^{9}+x^{14}y^{2}$ &
$y^{6}+x^{10}y^{8}+x^{14}y^{5}$ &
$x^{5}y^{8}+x^{11}y^{9}$ &
$x^{9}+x^{15}y^{3}$ &
24 & Fail & 34 & 42 & \textbf{24} \\

$(336,28,6)$ &
$x^{10}y+x^{12}y^{5}+x^{22}$ &
$x^{7}+x^{19}y+x^{26}y^{5}$ &
$x^{15}y^{2}+x^{25}y^{3}$ &
$x^{5}y^{3}+x^{21}y^{4}$ &
20 & Fail & 30 & 28 & \textbf{20} \\

$(336,28,6)$ &
$x^{12}+x^{16}y+x^{21}y^{2}$ &
$x^{8}y^{5}+x^{11}+x^{27}y$ &
$x^{12}+x^{25}y^{3}$ &
$x^{11}y^{3}+x^{24}y^{3}$ &
24 & Fail & 34 & 28 & \textbf{24} \\

$(216,18,6)$ &
$x^{4}y+x^{5}+x^{7}y+x^{15}y$ &
$xy+x^{9}y^{2}+x^{9}y^{3}+x^{13}y^{2}$ &
-- & -- &
18 & 26 & 26 & 26 & \textbf{18} \\

$(308,22,7)$ &
$x^{9}y+x^{9}y^{4}+x^{12}y^{2}$ &
$x^{10}y+x^{16}y^{6}+x^{18}y^{5}$ &
-- & -- &
20 & \textbf{20} & 28 & \textbf{20} & \textbf{20} \\

$(250,25,5)$ &
$x^{3}+x^{6}y^{3}+x^{9}y^{3}+x^{13}y^{2}$ &
$x^{6}y^{3}+x^{10}y+x^{22}y+x^{22}y^{2}$ &
-- & -- &
25 & 31 & 33 & 31 & \textbf{25} \\

$(360,6,30)$ &
$y^{4}+x^{3}y^{6}+x^{4}$ &
$y^{8}+y^{19}+x^{3}$ &
-- & -- &
24 & 26 & 32 & 32 & \textbf{24} \\

$(324,18,9)$ &
$x^{5}y^{5}+x^{16}y^{7}+x^{17}y$ &
$x^{8}y^{3}+x^{10}y^{3}+x^{11}y^{7}$ &
$x^{11}y^{3}$ &
$x^{8}y^{3}+x^{13}y^{7}$ &
20 & Fail & 28 & 36 & \textbf{20} \\

$(294,21,7)$ &
$x^{5}+x^{7}y^{5}+x^{9}y^{5}$ &
$x^{3}y^{4}+x^{7}y^{5}+x^{11}y^{4}$ &
$x^{5}+x^{11}y^{3}$ &
$x^{5}y$ &
22 & Fail & 30 & 28 & \textbf{22} \\

$(350,25,7)$ &
$x^{4}y+x^{10}y^{4}+x^{17}y^{3}$ &
$x^{8}y^{4}+x^{18}y^{3}+x^{19}y$ &
$x+x^{23}y^{2}$ &
$x^{6}y+x^{21}y^{4}$ &
24 & Fail & 32 & 42 & \textbf{24} \\

$(350,25,7)$ &
$x^{7}y^{6}+x^{15}+x^{18}y^{2}$ &
$x^{5}y^{2}+x^{18}y^{3}+x^{18}y^{5}$ &
$x^{23}y+x^{24}y^{5}$ &
$y^{3}+x^{11}y^{5}$ &
22 & Fail & 30 & 36 & \textbf{22} \\

$(312,26,6)$ &
$x^{7}y+x^{7}y^{2}+x^{24}y^{3}$ &
$x^{5}y^{4}+x^{10}y^{5}+x^{17}$ &
$x^{14}y^{3}+x^{17}y^{2}$ &
$x^{12}+x^{15}y^{5}$ &
24 & Fail & 32 & 34 & \textbf{24} \\

\bottomrule
\end{tabular}%
}
\end{table}
\subsection{Ablation Insights}
\vspace{-5pt}
\label{app:ablation_insights}
We evaluate each ablation using several random seeds. \autoref{fig:ablation} reports the validation negative log-likelihood (NLL) as a function of the training epoch. The bold curves represent the complete configuration, which includes training augmentation, shift canonicalization, and FiLM conditioning, whereas the lighter curves represent variants in which the component indicated in each panel is removed. The Tanner spectral branch enables the model to reach a lower validation NLL, demonstrating the benefit of incorporating global spectral information. Training augmentation has the strongest regularizing effect: without it, both the polynomial-only and dual-branch models overfit rapidly and fail to reach their best attainable validation performance. Shift canonicalization and FiLM conditioning also provide consistent improvements, particularly for the polynomial-only model, where global spectral information is unavailable.

\newcommand{\dsTotal}{58,704}
\newcommand{\dsBB}{49,243}
\newcommand{\dsPBB}{9,461}
\newcommand{\dsBBpct}{83.9}
\newcommand{\dsPBBpct}{16.1}
\newcommand{\dsOrbits}{55,271}
\newcommand{\dsCanon}{58,704}
\newcommand{\dsNmin}{18}
\newcommand{\dsNmax}{360}
\newcommand{\dsNmed}{216}
\newcommand{\dsLmin}{3}
\newcommand{\dsLmax}{30}
\newcommand{\dsMmin}{3}
\newcommand{\dsMmax}{30}
\newcommand{\dsKmin}{2}
\newcommand{\dsKmax}{216}
\newcommand{\dsKmed}{4}
\newcommand{\dsRatemin}{0.006}
\newcommand{\dsRatemax}{0.750}
\newcommand{\dsWAmin}{2}
\newcommand{\dsWAmax}{4}
\newcommand{\dsWBmin}{2}
\newcommand{\dsWBmax}{4}
\newcommand{\dsDmin}{2}
\newcommand{\dsDmax}{25}
\newcommand{\dsDmed}{12}
\newcommand{\dsDmean}{11.36}
\newcommand{\dsExact}{21,698}
\newcommand{\dsExactpct}{37.0}
\newcommand{\dsBound}{37,006}
\newcommand{\dsBoundpct}{63.0}
\newcommand{\dsCensored}{37,006}
\newcommand{\dsCensoredpct}{63.04}
\newcommand{\dsConsensus}{27,881}
\newcommand{\dsConsensuspct}{47.5}
\newcommand{\dsDisagree}{989}
\newcommand{\dsDisagreepct}{1.7}
\newcommand{\dsComplete}{28,870}
\newcommand{\dsCompletepct}{49.2}
\newcommand{\dsConflict}{0}
\newcommand{\dsMilp}{47,017}
\newcommand{\dsFallback}{37,006}
\newcommand{\dsFallbackpct}{63.0}
\newcommand{\dsMilpCert}{12,869}
\newcommand{\dsMilpCertpct}{21.9}
\newcommand{\dsCpuHours}{22,769}
\newcommand{\dsbposdHours}{344}
\newcommand{\dsbposdTrials}{5.3\!\times\!10^{7}}
\newcommand{\dsqdistevolHours}{3,408}
\newcommand{\dsqdistevolTrials}{8.0\!\times\!10^{9}}
\newcommand{\dsqdistrndmwHours}{1,317}
\newcommand{\dsqdistrndmwTrials}{2.1\!\times\!10^{9}}
\newcommand{\dsMilpHours}{17,700}
\newcommand{\dsSrcBpOsd}{20,841}
\newcommand{\dsSrcMilp}{12,982}
\newcommand{\dsSrcQdistEvol}{11,620}
\newcommand{\dsSrcLogicalBasisBound}{11,380}
\newcommand{\dsSrcQdistRndMw}{1,025}
\newcommand{\dsSrcWeightSweep}{856}

\newpage
\section{Dataset statistics}
\label{sec:dataset_stats}
\vspace{-3pt}

\begin{figure}[!h]
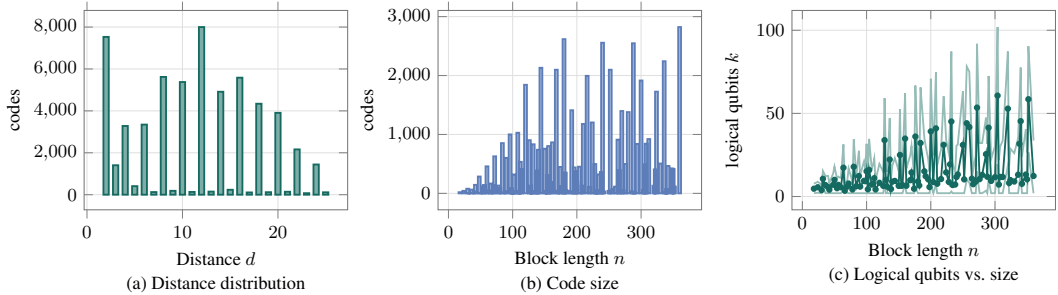

\centering
\input{tex_figures/dataset_dist.tex}%
\input{tex_figures/dataset_size.tex}%
\input{tex_figures/dataset_nk.tex}%
\definecolor{Ared} {RGB}{198, 60, 52}%
\definecolor{Bblue}{RGB}{ 52, 92,168}%
\definecolor{specc}{RGB}{ 20,106, 96}%
\definecolor{dark} {RGB}{ 60, 60, 60}%
\definecolor{greyc}{RGB}{150,150,150}%
\resizebox{\textwidth}{!}{%
\begin{tikzpicture}
\begin{groupplot}[
  group style={group size=3 by 1, horizontal sep=1.55cm},
  width=6.1cm, height=5.0cm,
  tick label style={font=\footnotesize},
  xlabel style={font=\small, align=center}, ylabel style={font=\small},
  grid=major, grid style={greyc!30, line width=0.3pt},
  axis line style={dark!70, line width=0.5pt},
  every axis plot/.append style={line width=0.9pt},
]

\nextgroupplot[xlabel={Distance $d$\\[2pt](a) Distance distribution},
  ylabel={codes}, ybar, bar width=2.6pt]
\addplot[fill=specc!55, draw=specc] coordinates {\dsDistAll};

\nextgroupplot[xlabel={Block length $n$\\[2pt](b) Code size},
  ylabel={codes}, ybar, bar width=1.4pt]
\addplot[fill=Bblue!45, draw=Bblue!80] coordinates {\dsSizeHist};

\nextgroupplot[xlabel={Block length $n$\\[2pt](c) Logical qubits vs.\ size},
  ylabel={logical qubits $k$}]
\addplot[specc!45, forget plot] coordinates {\dsNKlo};
\addplot[specc!45, forget plot] coordinates {\dsNKhi};
\addplot[specc, mark=*, mark size=1.1pt] coordinates {\dsNKmean};

\end{groupplot}
\end{tikzpicture}%
}
\caption{Corpus statistics. (a) distribution of the code distance;
(b) distribution of block length $n=2\ell m$; (c) mean logical dimension $k$
per block length, with $\pm$ one standard deviation. Throughout, the distance of a code
without an exact solution is its best available upper bound.}
\label{fig:dataset}
\end{figure}

Figure~\ref{fig:dataset} presents the dataset's distributions. The dataset holds 58,704 codes, 83.9\% of them
CSS BB and 16.1\% non-CSS PBB extensions, spanning block lengths $n=18$ to $360$, lattices $\ell,m\in[3,30]$,
$k=2$ to $216$ logical qubits and rates $k/n$ between 0.006 and
0.750. All heuristics used to create the upper bound distance for this dataset were implemented by the codeDistance package \citep{webster2026distance}.

\paragraph{Distributions.}
The distance distribution in Figure~\ref{fig:dataset}(a) is strongly right-skewed: the median is $d=12$ against a maximum of $25$. The distance of a code is defined to be its best available upper bound. Figure~\ref{fig:dataset}(b) describes the codes' length $n$ distribution. Figure~\ref{fig:dataset}(c) shows the logical qubits in codes as a function of the block length.

\section{Feature Construction and Embedding}
\label{sec:feature_embedding}

The two BB-Net branches operate on different raw feature spaces. Polynomial tokens combine a learned polynomial-type representation with cyclic Fourier coordinates, whereas Tanner spectral tokens contain explicitly computed character-response features. Each raw token is projected to the dimension of $d_{\mathrm{model}}$ before self-attention. The following is a description of the full feature computation on the spectral branch:

\paragraph{Notation.}
For integers $p,q$ we write $\gcd(p,q)$ for their greatest common divisor, the largest integer dividing both, and $\operatorname{lcm}(p,q)=pq/\gcd(p,q)$ for their least common multiple, the smallest positive integer that both divide. For a complex number $z$, $\overline{z}$ is its complex conjugate, $|z|$ its modulus, and $\operatorname{Re}(z),\operatorname{Im}(z)$ its real and imaginary parts. The symbol $\mathbf{1}[\,\cdot\,]$ denotes the indicator function, equal to $1$ when the enclosed condition holds and $0$ otherwise; $p\mid q$ reads ``$p$ divides $q$''. Finally, $\mathbin{\|}$ denotes concatenation of vectors.

\paragraph{Character responses.}
For each character $\chi=\chi_{u,v}$, define the response of polynomial $P$ as
\begin{equation}
\begin{aligned}
z_P(\chi)=\widehat{P}(\chi)=\sum_{(a,b)\in S_P}
\exp\left[
-2\pi i
\left(
\frac{ua}{\ell}+\frac{vb}{m}
\right)
\right],
\qquad
w_P=|S_P|,
\end{aligned}
\end{equation}
for $P\in\{A,B,C,D\}$, where $w_P$ is the number of monomials in $P$.

Its normalized magnitude is
\begin{equation}
a_P(\chi)
=
\begin{cases}
|z_P(\chi)|/w_P, & w_P>0,\\
0, & w_P=0,
\end{cases}
\end{equation}
so that $0\leq a_P(\chi)\leq 1$.
We also compute the joint energy, cross-spectrum, and relative spectral difference:
\begin{align}
s_2(\chi)=\frac{|z_A(\chi)|^2+|z_B(\chi)|^2}{w_A^2+w_B^2},\ \ q_{AB}(\chi)=\frac{z_A(\chi)\overline{z_B(\chi)}}{w_Aw_B}, \ \ \delta_{AB}(\chi)=\frac{|z_A(\chi)-z_B(\chi)|}{w_A+w_B}.
\end{align}
All three are normalized by polynomial weight so that they remain comparable across codes with different numbers of monomials.

\paragraph{Character order.}
The order of $\chi_{u,v}$ is
\begin{equation}
\begin{aligned}
o_{\ell}(u)=\frac{\ell}{\gcd(u,\ell)},\ \ 
o_m(v)=\frac{m}{\gcd(v,m)}, \ \ 
o_{\chi}=\operatorname{lcm}\bigl(o_{\ell}(u),o_m(v)\bigr).
\end{aligned}
\end{equation}
The feature $\mathbf{1}[2\mid o_{\chi}]$ equals $1$ when $o_{\chi}$ is even. The feature $\mathbf{1}[(u,v)=(0,0)]$ flags the trivial character, at which $z_P(\chi)=w_P$ for every $P$.

\paragraph{Spectral feature vector.}
For readability, we specify the 18-dimensional spectral feature vector in three groups. All response-dependent quantities below are evaluated at $\chi$.

The first group collects the per-polynomial magnitudes and the pairwise response statistics:
\begin{equation}
\begin{aligned}
\mathbf{s}^{(1)}_{\chi}
=\bigl[&
a_A,\ a_B,\ a_C,\ a_D,\ a_A^2,\ a_B^2,\ s_2,\ \operatorname{Re}(q_{AB}),\ \operatorname{Im}(q_{AB}),\ |q_{AB}|
\bigr]
\in\mathbb{R}^{10}.
\end{aligned}
\end{equation}

The second group sharpens the near-degenerate regime:
\begin{equation}
\begin{aligned}
\mathbf{s}^{(2)}_{\chi}
=\bigl[\delta_{AB},\ e^{-8s_2},\ e^{-8(1-\min(a_A,a_B))}
\bigr]
\in\mathbb{R}^{3},
\end{aligned}
\end{equation}

The third group contains invariants of the character itself, independent of the polynomial supports:
\begin{equation}
\begin{aligned}
\mathbf{s}^{(3)}_{\chi}
=\biggl[&
\frac{o_{\chi}}{\ell m},\
\frac{\log(1+o_{\chi})}{\log(1+\ell m)},\ \frac{1}{o_{\chi}},\
\mathbf{1}[2\mid o_{\chi}],\ \mathbf{1}[(u,v)=(0,0)]
\biggr]
\in\mathbb{R}^{5}.
\end{aligned}
\end{equation}

The complete feature vector is
\begin{equation}
\mathbf{s}_{\chi}
=
\mathbf{s}^{(1)}_{\chi}
\mathbin{\|}
\mathbf{s}^{(2)}_{\chi}
\mathbin{\|}
\mathbf{s}^{(3)}_{\chi}
\in\mathbb{R}^{18}.
\label{eq:spectral_features}
\end{equation}

\end{document}